\documentclass{amsart}
\usepackage[all]{xy}

\newtheorem{theorem}{Theorem}[section]
\newtheorem{lemma}{Lemma}[section]
\newtheorem{proposition}{Proposition}[section]
\newtheorem{corollary}{Corollary}[section]

\newtheorem{example}{Example}[section]

\theoremstyle{remark}
\newtheorem{remark}{Remark}[section]
\numberwithin{equation}{section}

\newcommand{\internalcomment}[1]{}

\begin{document}

\title[application in inverse problems]{Application of
  approximation theory by nonlinear manifolds in Sturm-Liouville
  inverse problems}
\author{Amadeo Irigoyen}
\address{Universit{\'e} Paris VI - Pierre et Marie Curie, UMR 7586, 175,
  rue du Chevaleret 75013 Paris}

\email{
\begin{minipage}[t]{5cm}
amadeo@math.jussieu.fr
\end{minipage}
}

\begin{abstract}

We give here some negative results in Sturm-Liouville inverse theory,
meaning that we cannot approach any of the potentials with $m+1$
integrable derivatives on $\mathbb{R}^+$ by an $\omega$-parametric
analytic family better than order of
$(\omega\ln\omega)^{-(m+1)}$.

Next, we prove an
estimation of the eigenvalues and characteristic values of a
Sturm-Liouville operator and some properties of the solution of a certain
integral equation. This allows us to deduce
from~\cite{henkin} some positive results about the best reconstruction
formula by giving an almost optimal formula of order of
$\omega^{-m}$. 

\end{abstract}

\maketitle

\tableofcontents

\section{Introduction}

In this paper we deal with one of the problems of Sturm-Liouville inverse
theory. We consider the equation on the half-axis $\mathbb{R}^+$
$$-\frac{d^2y}{dx^2}-\omega^2Qy=\lambda y\,,$$
where the potential $-\omega^2Q$ is strictly negative with many
derivatives which are locally integrable and polynomially decreasing
at infinity, and $\omega$ a large enough parameter. We know
(see~\cite{calogero}) that the
operator $-\frac{d^2}{dx^2}-\omega^2Q$ admits $N(\omega)$ (of
order $\omega$) negative
eigenvalues $-\xi_j^2$ and for each one of these an unique
eigenfunction $\phi_j$ such that
$$\phi_j(0)=0\;\text{ and }\int_0^{\infty}|\phi_j(x)|^2dx=1\;.$$
We put $C_j=(\phi'_j(0))^2$, which is called the characteristic value
associated to $\xi_j$.

The problem deals with the reconstruction of the potential
$-\omega^2Q$ from given data of the eigenfunctions, and more
precisely from the eigenvalues $\xi_j$ and characteristic values
$C_j$. The motivation comes from one side, of the questions of
seismology (see~\cite{henkin},~\cite{henkin2}), and on the other side
from the results of Lax and Levermore on the small dispersion limit of
the KdV equation (see~\cite{lax-levermore}). We know that there are
more or less explicit approximation formulas coming from the works of
Gelfand, Levitan, Kohn and Jost
(see~\cite{gelfand},~\cite{levitan}). Moreover, motivated by Lax
and Levermore, G. Henkin and N. Novikova gave in~\cite{henkin} some
results with more precision on the convergence~: more precisely, they
gave formulas that approximate any $Q$ with $m+1$ locally
integrable derivatives uniformly
on any $[0,X]$ when $\omega\rightarrow+\infty$, at least at order
$\frac{1}{\omega^m}$.

Then there is a natural question (see~\cite{henkin}, p. 22) about a better
approximation~: is there another formula (and  can it in this case
be made explicit) which gives a faster convergence to $Q$ ?
\bigskip

There are two cases in which we give an answer : in the first
with $m=1$ there is an explicit approximation formula of
Gelfand-Levitan type :
$$Q_{\omega}^0(x)=\frac{2}{\omega^2}\frac{d^2}{dx^2}\ln|\det(W_{s,r})(x)|,$$
with
$$W_{s,r}(x)=\frac{2sh(\xi_r+\xi_s)x}{\xi_r+\xi_s}-(1-\delta_{s,r})
\frac{2sh(\xi_r-\xi_s)x}{\xi_r-\xi_s}-\delta_{s,r}\left(2x-
\frac{4\xi_j^2}{C_j}\right).$$
This formula (see~\cite{henkin}) approximates any $Q$ by primitives
uniformly on any 
$[0,X]$, with order $\frac{\ln\omega}{\sqrt{\omega}}$ (but hopefully
the precision could be better, like $\frac{1}{\omega^3}$). We then have the
following result :

\begin{theorem}\label{pthm1}

Let $\mathcal{Q}_2$ be the class of functions $Q$, defined on
$\mathbb{R}^+$, strictly positive, polynomially decreasing with $2$
locally integrable derivatives such that $Q'(0)=0$, and with operators
$-\frac{d^2}{dx^2}-\omega^2Q$. For all $N=O(\omega)$, let
$\psi(x,\zeta)$ be a function which is defined on
$\mathbb{R}^+\times\mathbb{C}^N$, of class $C^1$ with
respect to $x$ and entire
of exponential type with respect to
$\zeta\in\mathbb{C}^N$. Then the approximation of
$$\int_0^{\cdot}\mathcal{Q}_2:=\left\{\left(x\mapsto\int_0^xQ(t)dt
\right),\;Q\in\mathcal{Q}_2\right\}$$
uniformly on any $[0,X]$, by the family
$$\left\{\left(x\mapsto\left(\frac{1}{\psi}\frac{\partial\psi}{\partial x}
\right)(x,\zeta)\right), \,\zeta_j=O(\omega^r),
\,\forall\,j=1,\ldots,N\right\},$$
when $\omega\rightarrow\infty$, cannot be better than order of
$$\frac{1}{(\omega\ln\omega)^3}\,.$$

Moreover there is a case where the approximation is (at least) of order of
$\frac{\ln\omega}{\sqrt{\omega}}$, which is given by the formula of
Gelfand-Levitan type :
$$\Psi(x,\zeta)=\det\widetilde{W}_{s,r}(x,\zeta),$$
with
$$\widetilde{W}_{s,r}(x,\zeta)=\frac{2sh(\zeta_r+\zeta_s)x}{\zeta_r+\zeta_s}
-(1-\delta_{s,r})\frac{2sh(\zeta_s-\zeta_r)x}{\zeta_s-\zeta_r}
-\delta_{s,r}\left(2x-\exp\left(\zeta_{r+N(\omega)}\right)\right),$$
$s,\,r=1,\ldots,N(\omega)$, where $N(\omega)$ is the number of
eigenvalues $\xi_j$ and characteristic values $C_j$ of the operator
$-\frac{d^2}{dx^2}-\omega^2Q$, whose optimizing element can be
chosen to be
$$\zeta_j(Q)=\xi_j(Q)\text{ and }\zeta_{j+N(\omega)}(Q)
=\ln\frac{4\xi_j^2(Q)}{C_j(Q)},\,j=1,\ldots,N(\omega).$$

\end{theorem}

This result will be proved in section~\ref{application} : the
Gelfand-Levitan type formula can indeed be seen as a nonlinear
family with respect to the parameters $\xi_j$ and $C_j$. In general one
approximation formula can be considered as an approximation of any
compact subset of $\mathcal{Q}_2$ by a nonlinear manifold. More precisely we
deal with the case when the family is represented by an entire
function of exponential type. Thanks to the negative results on
approximation theory (which will be given in section~\ref{rappel},
see~\cite{amadeo}), we will be able to get lower bounds for such
approximation formulas. Notice (see section~\ref{rappel}) that the
decreasing condition of any $Q$ is not usefull to prove the negative case.
However an estimation of the
eigenvalues and characteristic values (which will be proved on
section~\ref{estimation}) is necessary~:

\begin{proposition}

Let $-\omega^2Q$ be a strictly negative and integrable potential of
class $C^1$ with $Q'(0)=0$, and that is polynomially decreasing. Then for
any sufficiently large $\omega$ and $j=1,\ldots,N(\omega)$, one has
$$\frac{1}{a\omega^b}\leq\xi_j\leq a\omega^b\text{ and }
\frac{1}{\alpha\exp\left(\beta\omega^{\gamma}\right)}\leq
\frac{4\xi_j^2}{C_j}\leq\alpha \exp\left(\beta\omega^{\gamma}\right).$$

\end{proposition}

This decreasing condition of $Q$ is assumed in order to apply rigorously
the WKB method but apparently it should still be true in a more
general case (see~\cite{lax-levermore}, III).
\bigskip

In order to get a better estimation, not only for $2$ derivatives but
for $m+1$ derivatives (see~\cite{henkin}), we have to deal with not
explicit formulas (because any better estimate is not achieved by this
explicit formula given above) which approximate $Q$ 
with order (at least) of $\frac{1}{\omega^m}$ (we think that the
precision could be better,
of order of $\frac{1}{\omega^{m+1}}$ ). However these formulas
have, in the case where the derivatives of $Q$ vanish at $0$, the following
expression~:
$$Q_{\omega}(x)=\frac{2}{\omega^2}\left(-\frac{d}{dx}A(x,x)+\frac{d^2}{dx^2}
\det T_{j,k}(x)\right)\,,$$
where $A(x,y)$ is the solution of the integral equation
$$A(x,y)+\int_0^xA(x,s)\Phi(s,y)ds+\Phi(x,y)=0$$
with
$$\Phi(x,y)=\int_0^{\infty}\frac{\sin{kx}}{k}\frac{\sin{ky}}{k}
\left(\sqrt{k^2+\omega^2Q(0)}-k\right)\frac{2k}{\pi}dk,$$
and for all $j,k=1,\ldots,N(\omega)$
\begin{eqnarray*}
T_{j,k}(x) & = & \frac{4\xi_j^2}{C_j}\delta_{j,k}\\
& + & 4\int_0^x\left(sh(\xi_jt)+\int_0^tA(t,s)sh(\xi_js)ds\right)
\left(sh(\xi_kt)+\int_0^tA(t,s)sh(\xi_ks)ds\right)\,dt\,.
\end{eqnarray*}
Even better, these formulas can be seen as analytic families with
respect to the parameters
$\xi(Q)=\left(\xi_1,\ldots,\xi_{N(\omega)}\right),
\;C(Q)=\left(C_1,\ldots,C_{N(\omega)}\right)$
and $\omega^2Q(0)$. Another application of our negative results (and a
positive result given in~\cite{henkin}) gives the following theorem :

\begin{theorem}\label{pthm2}

Consider $\mathcal{Q}_{m+1}$ the class of strictly positive functions
$Q$, strictly decreasing with $m+1$ locally integrable derivatives
which vanish at $0$,
and their Sturm-Liouville operators $-\frac{d^2}{dx^2}-\omega^2Q$.
For any $N,M=O(\omega)$, let
$\psi(x,\zeta,w)$ (resp. $k(x,\zeta,w)$ ) be a function defined on
$\mathbb{R}\times\mathbb{C}^N\times\{\Re e\,z>0\}^M$, of class $C^2$
(resp. continuous) with respect to $x\in\mathbb{R}^+$ and holomorphic with
exponential kind with respect to
$(\zeta,w)\in\mathbb{C}^N\times\{\Re e\,z>0\}^M$.

Then the approximation of $\mathcal{Q}_{m+1}$
uniformly on any $[0,X]$ by the family
$$\left\{\left(x\mapsto k(x,\zeta,w)+
\frac{\partial}{\partial x}
\left(\frac{1}{\psi}\frac{\partial\psi}{\partial x}\right)(x,\zeta,w)
\right),\;\zeta_j=O(\omega^r),\,
|w_i-\omega^{r_i}|<\omega^{r_i}\right\},$$
when $\omega\rightarrow\infty$, cannot be better than of order of
$$\frac{1}{(\omega\ln\omega)^{m+1}}\,.$$
In addition there is an almost optimal approximation formula defined as
$$k(x,w)=-\frac{\partial}{\partial x}A(x,x,w)
\text{ and }\Psi(x,\zeta,w)=\det T_{j,k}(x,\zeta,w)$$
for $(\zeta,w)\in\mathbb{C}^{2N(\omega)}\times\{\Re e\,z>0\}$, which gives a
positive result of order of
$$\frac{1}{\omega^m}.$$
Moreover $Q\in\mathcal{Q}$ being given, such an element
$(\zeta(Q),w(Q))$ can be chosen as
$$\zeta_j(Q)=\xi_j(Q),\;\zeta_{j+N(\omega)}(Q)
=\ln\frac{4\xi_j^2(Q)}{C_j(Q)},
\;j=1,\ldots,N(\omega),\text{ and }\;w(Q)=\omega^2Q(0).$$

\end{theorem}

As in the previous case, the decreasing condition is not usefull to
prove the negative case.
However, in order to associate negative and positive results we must
prove some properties of the solution $A(x,y)$ (see
section~\ref{solequint}) :

\begin{proposition}

The solution $A\left(x,y,\omega^2Q(0)\right)$ exists and is unique,
and can be holomorphically extended by $A(x,y,w)$ on the half-plane 
$W=\{\Re e\,w>0\}$ with the following properties : 
$$x\in\mathbb{R}^+\mapsto A(x,y,w)\in L^2_y([0,x]),$$
is continuously differentiable and of polynomial kind with respect to
$w\in W$.

\end{proposition}

We finish then by giving some examples of inverse problems with some
possible analogous applications.
\bigskip

I would like to thank G. Henkin for interesting problems and
improving discussions about this work.
\bigskip

\section{Some results on approximation theory by nonlinear
  manifolds}\label{rappel}

We consider the compact set
$\Lambda_l(I^s),\;I=[0,1],\;s\in\mathbb{N}^{\ast},\;l>0$,
$$\Lambda_{l,s}=\Lambda_l(I^s)=\left\{f\in C^l(I^s),\;\forall\,j,\;
0\leq j\leq m,\;\left\|f^{(j)}\right\|_{\infty}\leq1,\;
\left\|f^{(m)}\right\|_{\infty}\leq1\right\}\,,$$
where $l=m+\alpha,\;m\in\mathbb{N},\;0<\alpha\leq1$ ($m=-[l]-1$), and
$$\|f\|_{\alpha}=\sup_{x\neq y}\frac{|f(x)-f(y)|}{\|x-y\|^{\alpha}}\,,$$
with $\|\cdot\|$ the usual euclidian norm ($\Lambda_{l,s}$ is a
compact subset of $\left(C^0(I^s),\|\cdot\|_{\infty}\right)$ and
$\left(L^1(I^s),\|\cdot\|_1\right)$ ).

We remind the theorem of Vitushkin
(see~\cite{vitushkin},~\cite{warren}) :

\begin{theorem}\label{vitushkin}

Let consider the family
$$P_{n,d}=\left\{P(\zeta)=\sum_{|k|\leq d}c_k(x)\zeta^k,\;
\zeta=(\zeta_1,\ldots,\zeta_n)\in\mathbb{R}^n\right\}\,,$$
where $n\geq1,\;d\geq2$ are integers, and $c_k\in C(I^s)$
(resp. $L^1(I^s)$).

Then $\exists\,h\in\Lambda_{l,s}$, such that
$\forall\,\zeta\in\mathbb{R}^n$,
$$\|h-P(\zeta)\|\geq\frac{C(l,s)}{(n\log d)^{\frac{l}{s}}}\,,$$
where $\|\cdot\|$ is the uniform norm $\|\cdot\|_{\infty}$
(resp. $\|\cdot\|_{L^1}$), $C(l,s)=C_{\infty}(l,s)$
(resp. $C_{L^1}(l,s)$).

Equivalently, if $\mathcal{P}_{n,d}$ is the set of families of $C(I^s)$
(resp. $L^1(I^s)$) which are parametrized by $n$ variables and
polynomially of degree (at most) $d$, then
$$D_{n,d}(\Lambda_{l,s}):=\inf_{P\in\mathcal{P}_{n,d}}\,
\sup_{h\in\Lambda_{l,s}}\,\inf_{\zeta\in\mathbb{R}^n}\|h-P(\zeta)\|
\geq\frac{C(l,s)}{(n\log d)^{\frac{l}{s}}}\,.$$

The complete proof of this theorem with
precision of constants is given in~\cite{amadeo} :
$$C_{\infty}(l,s)=\frac{1}{\sqrt{s}\,2^{l+1}8^{\frac{l}{s}}([l]+1)^{[l]+1}
(4(1+e))^{s([l]+1)}}\;,$$
and
$$C_{L^1}(l,s)=\frac{(([l]+1)!)^{2s}}{5\sqrt{s}\,2^{l+2}18^{\frac{l}{s}}
([l]+1)^{[l]+1}((2[l]+3)!)^s(1+e)^{s([l]+1)}}\;.$$

\end{theorem}
\bigskip

\begin{remark}\label{etoile}

In the continuous case when $s=1$ and $l>1$, we can even assume that 
$h\in\Lambda_l\cap C^l(I^s)$ and satisfies (see corollary 3
in~\cite{amadeo}) :
$$(\ast)
\begin{cases}
\frac{C_{\infty}(l,1)}{(n\log d)^{\frac{l}{s}}}\leq\|h\|_{\infty}=
|h(x(\zeta))|\leq\frac{2^lC_{\infty}(l,1)}{(n\log d)^{\frac{l}{s}}},\\
h(x(\zeta),\zeta)P(x(\zeta),\zeta)\leq0\\
h(x)=0,\;\forall\,x\in\left[0,\frac{1}{20}\right].
\end{cases}
$$

\end{remark}

We remind the following theorem as well (the proof is given
in~\cite{amadeo}), which is an analogous result of the
theorem~\ref{vitushkin} for the analytic case :

\begin{theorem}\label{th1bis}

Let consider the domain $W=\{\Re e\,z>0\}$ and for $N\geq2,\;M\geq0$,
$$f(x,\zeta,w)=f(x_1,\ldots,x_s,\zeta_1,\ldots,\zeta_N,w_1,\ldots,w_M),\,
x\in{I^s},\,\zeta\in\mathbb{R}^N,\,w\in\left(\mathbb{R}^+\right)^M$$
where $f$ is entire with respect to $\zeta\in\mathbb{C}^N$,
holomorphic with respect to $w\in W^M$ and continuous
with respect to $x\in I^s$. Assume that
$\forall\,(\zeta,w)\in\mathbb{C}^N\times W^M$,
$$\|f(\cdot,\zeta,w)\|_{\infty}\leq Ae^{u(N+M)^v}
e^{b(N+M)^t\left(\|\zeta\|_1^d+\|w\|_1^d\right)},$$
where $A,u,v,b,t,d\in[1,+\infty[$ and
$\|\zeta\|_1=|\zeta_1|+\ldots+|\zeta_N|$ (as well as $\|w\|_1$).

Consider the subset
$$\Omega_{N,M}=\left\{(\zeta,w)\in\mathbb{R}^N\times
\left(\mathbb{R}^+\right)^M,\;|\zeta_j|\leq B_1(M+N)^{r_1},\;
|w_i-a_i|\leq(1-\varepsilon)a_i\right\},$$
where for all $i$, $a_i\leq B_2(M+N)^{r_2}$, and
$B_1,\,B_2,\,r_1,\,r_2\geq1$ and $0<\varepsilon<1$. 
\bigskip

Then $\exists\,h\in\Lambda_{l,s}$ such that
$\forall\,(\zeta,w)\in\Omega_{N,M}$
$$\|h-f(\cdot,\zeta,w)\|_{\infty}\geq
\frac{C}{\left((M+N)\log_{[2]}(M+N)\right)^{\frac{l}{s}}}\,.$$

In the case $s=1$ and $l>1$, we can in addition assume that $h$ is
identically zero in $\left[0,\frac{1}{20}\right]$ (property $(\ast)$ )
and satisfies that $\forall\,(\zeta,w)\in\Omega_{N,M}$,
$\exists\,x_{\zeta,w}$ such that
$$(\ast')
\begin{cases}
\frac{C_{\infty}(l,1)}{((N+M)\log(K+K^2))^l}\leq|h(x_{\zeta,w})|
\leq\|h\|_{\infty}
\leq\frac{2^{l}C_{\infty}(l,1)}{((N+M)\log(K+K^2))^l},\\
h(x_{\zeta,w})P_K(x_{\zeta,w},\zeta,w)\leq0,\\
\left\|R_K(\cdot,\zeta,w)
\right\|_{\infty}\leq\frac{C_{\infty}(l,1)}{2((N+M)\log(K+K^2))^l}\,,
\end{cases}
$$
where $f(\cdot,\zeta,w)=P_K(\cdot,\zeta,w)+R_K(\cdot,\zeta,w)$ is the
decomposition of $f$ as a polynomial $P_K$ and the remainder $R_K$, with
$K=O\left((N+M)^{\alpha}\right)$.

\end{theorem}

Although it seems to be unnatural, the choice of the half-plane $W$ is
motived by the applications in the last section (see
theorem~\ref{optim2}) where we will choose $M=1$.
We deduce here the following corollaries from theorem~\ref{th1bis},
which will be usefull in the section~\ref{application}. 

\begin{corollary}\label{inverse1}

Let $\psi(x,\zeta,w)$ be a function, defined on
$[0,1]\times\mathbb{C}^N\times W^M$, of class
$C^2$ with respect to $x$, such that for all $x\in[0,1]$ the
functions $\psi(x,\zeta,w)$, $\frac{\partial\psi}{\partial x}(x,\zeta,w)$ 
and $\frac{\partial^2\psi}{\partial x^2}(x,\zeta,w)$ satisfy the
conditions of theorem~\ref{th1bis}. Let be in the
other side $k(x,\zeta,w)$ continue with respect to $x$ and analytic of
exponential type with respect to $(\zeta,w)$ such that the restriction on
$\mathbb{R}^N\times\left(\mathbb{R}^+\right)^M$ is of polynomial type.

Moreover assume that
$\forall\,(\zeta,w)\in\Omega_{N,M}$
$$\frac{1}{\psi(0,\zeta,w)}=O\left(e^{\alpha(M+N)^{\beta}}\right)$$
and
$$\frac{\partial\psi}{\partial x}(0,\zeta,w)=0.$$
Let finally $b_{M,N}>0$ be a constant such that $b_{M,N}$ and
$\frac{1}{b_{M,N}}$ are of polynomial type at $M+N$.
\bigskip

Then the approximation of $\Lambda_l([0,1])$ by the family 
$$\left\{\left(x\in[0,1]\mapsto\frac{1}{b_{M,N}}\left(k(x,\zeta,w)+
\frac{\partial}{\partial x}
\left(\frac{1}{\psi}\frac{\partial\psi}{\partial x}\right)(x,\zeta,w)\right)
\right),\,(\zeta,w)\in\Omega_{N,M}\right\},$$
in the uniform sense on $[0,1]$, when $N+M\rightarrow+\infty$,
cannot be better than 
$$\frac{\widetilde{C}}{(N+M)^l(\ln(N+M))^l}.$$

In addition, a function $h$ which satisfies the minoration can be
choosen in $\Lambda_l\cap C^{[l]}([0,1])$ and be identically zero on
$\left[0,\frac{1}{20}\right]$.

\end{corollary}

\begin{proof}

We can assume that $\forall\,(\zeta,w)$, $\psi(0,\zeta,w)>0$ :
indeed, the condition about $\psi(0,\zeta,w)$ shows that it nevers
vanishes on the (connected) set
$\mathbb{R}^N\times\left(\mathbb{R}^+\right)^m$, 
then has the same sign  and can be replaced by $-\psi$ (this does not
change the family).
\bigskip

Let $\psi,\,k$ be given and set
$$\widetilde{\psi}(x,\zeta,w)=e^{\widetilde{k}(x,\zeta,w)}
\psi(x,\zeta,w)\,,$$ 
with
$\widetilde{k}(x,\zeta,w)=\int_0^xdt\int_0^tk(s,\zeta,w)ds$. First we
claim that the theorem~\ref{th1bis} is still true with
$\widetilde{\psi}$ (although it can be not of exponential type) :
there is indeed
$$\left\|e^{\widetilde{k}(\cdot,\zeta,w)}\right\|_{\infty}
\leq\widetilde{A}e^{\widetilde{\alpha}(N+M)^{\widetilde{\beta}}},$$
then by applying the theorem~\ref{th1bis} to
$\widetilde{A}e^{\widetilde{\alpha}(N+M)^{\widetilde{\beta}}}\psi$ we
get by property $(\ast')$ with $K=O(N+M)$ and for all
$(\zeta,w)\in\Omega_{N,M}$ :
$$\widetilde{A}e^{\widetilde{\alpha}(N+M)^{\widetilde{\beta}}}
\psi(x,\zeta,w)=P_K(x,\zeta,w)+R_K(x,\zeta,w)$$
and
$$\left\|\frac{e^{\widetilde{k}(\cdot,\zeta,w)}}
{\widetilde{A}e^{\widetilde{\alpha}(N+M)^{\widetilde{\beta}}}}
R_K(\cdot,\zeta,w)\right\|_{\infty}
\leq\frac{C_{\infty}(l,1)}{2((N+M)\log(K+K^2))^l}\,.$$
On the other side there is $h\in\Lambda_l([0,1])$ associate to the
polynomial $P_K$, such that
$\forall\,(\zeta,w),\;\exists\,x_{\zeta,w}$ wich satisfies
$$\frac{C_{\infty}(l,1)}{((N+M)\log(K+K^2))^l}\leq|h(x_{\zeta,w})|\;
\text{ and }h(x_{\zeta,w})P_K(x_{\zeta,w},\zeta,w)\leq0,$$
hence
\begin{eqnarray*}
\left\|h-\widetilde{\psi}(\cdot,\zeta,w)\right\|_{\infty} & \geq & 
\left|h(x_{\zeta,w})-\frac{e^{\widetilde{k}(x_{\zeta,w},\zeta,w)}}
{\widetilde{A}e^{\widetilde{\alpha}(N+M)^{\widetilde{\beta}}}}
P_K(x_{\zeta,w},\zeta,w)\right|-
\left\|\frac{e^{\widetilde{k}(\cdot,\zeta,w)}}
{\widetilde{A}e^{\widetilde{\alpha}(N+M)^{\widetilde{\beta}}}}
R_K(\cdot,\zeta,w)\right\|_{\infty}\\
& \geq & \frac{C_{\infty}(l,1)}{2((N+M)\log(K+K^2))^l}\,,
\end{eqnarray*}
and $K$ being polynomial at $(N+M)$, this proves theorem~\ref{th1bis}
with $\widetilde{\psi}$.
\bigskip

Now we can prove the corollary. Set
$$\widetilde{\chi}(x,\zeta_1,\ldots,\zeta_N,\zeta_{N+1},w)=e^{\zeta_{N+1}}
\left(\frac{\partial^2\widetilde{\psi}}{\partial^2x}\widetilde{\psi}-
\left(\frac{\partial\widetilde{\psi}}
{\partial x}\right)^2\right)(x,\zeta,w)\,,$$
and we see that
$$\widetilde{\chi}(\cdot,\zeta,w)
=e^{2\widetilde{k}(\cdot,\zeta,w)}\chi(\cdot,\zeta,w).$$
where $\chi$ still fulfills the conditions of theorem~\ref{th1bis}. It
follows from above that there is
$h\in\Lambda_l([0,1])$, such that for all $(\zeta,w)\in\Omega_{N,M}$,
$\exists\,x=x_{\zeta,\zeta_{N+1},w}$ which satisfies :
$$
\begin{cases}
\frac{C_{\infty}(l,1)}{((N+M+1)\log(K+K^2))^l}\leq|h(x)|\leq\|h\|_{\infty}
\leq\frac{2^lC_{\infty}(l,1)}{((N+M+1)\log(K+K^2))^l},\\
h(x)e^{2\widetilde{k}(x,\zeta,w)}P_K(x,\zeta,\zeta_{N+1},w)\leq0,\\
\left\|e^{2\widetilde{k}(\cdot,\zeta,w)}R_K(\cdot,\zeta,\zeta_{N+1},w)
\right\|_{\infty}\leq\frac{C_{\infty}(l,1)}{2((N+M+1)\log(K+K^2))^l}\,,
\end{cases}
$$
where $\chi=P_K+R_K$ and $K=O((N+M)^{\gamma})$.

Moreover $h\in C^{[l]}([0,1])$ and is identically zero on
$\left[0,\frac{1}{20}\right]$.
\bigskip

Next there is still for all $\mu\in[0,1]$,
$$h(x)\,\mu\,e^{2\widetilde{k}(x,\zeta,w)}P_K(x,\zeta,\zeta_{N+1},w)
\leq0$$
and
$$\left\|\mu\,e^{2\widetilde{k}(\cdot,\zeta,w)}
R_K(\cdot,\zeta,\zeta_{N+1},w)\right\|_{\infty}
\leq\frac{C_{\infty}(l,1)}{2((N+M+1)\log(K+K^2))^l},$$
then
$$\left|h(x)-\mu\widetilde{\chi}(x,\zeta,\zeta_{N+1},w)\right|
\geq\frac{C_{\infty}(l,1)}{2((N+M+1)\log(K+K^2))^l}\,.$$
Now $\psi$ is of exponential type (and $\widetilde{k}(0,\zeta,w)=0$)
then for all $(\zeta,w)\in\Omega_{N,M}$,
$$0<\widetilde{\psi}(0,\zeta,w)=\psi(0,\zeta,w)=
O\left(e^{\alpha'(N+M)^{\beta'}}\right),$$
as well for $\frac{1}{\widetilde{\psi}(0,\zeta,w)}$ and
$e^{b_{N,M}}$ from hypothesis, hence
$$\ln\left(\frac{e^{b_{N,M}}}{b_{N,M}\widetilde{\psi}(0,\zeta,w)^2}\right)
=O\left((N+M)^{r'}\right),$$ 
which can be chosen as value for the parameter $\zeta_{N+1}$ ;
there is for all $\mu\in[0,1]$ : 
$$\left|h(x)-\frac{\mu e^{b_{N,M}}}{b_{N,M}}
\frac{\frac{\partial^2\widetilde{\psi}}
{\partial^2x}(x,\zeta,w)\,\widetilde{\psi}(x,\zeta,w)
-\left(\frac{\partial\widetilde{\psi}}{\partial x}\right)^2(x,\zeta,w)}
{\left(\widetilde{\psi}(0,\zeta,w)\right)^2}\right|\geq
\frac{C_{\infty}(l,1)}{2((N+M+1)\log(K+K^2))^l}.$$
\bigskip

Now assume that
$e^{-b_{N,M}}\left(\frac{\widetilde{\psi}(0,\zeta,w)}{\widetilde{\psi}
(x,\zeta,w)}\right)^2\leq1$.
Then it can be substituted in the inequality as value for $\mu$ to get
:
\begin{eqnarray*}
\left|h(x)-\frac{1}{b_{N,M}}\frac{\frac{\partial^2\widetilde{\psi}}
{\partial^2x}(x,\zeta,w)\,\widetilde{\psi}(x,\zeta,w)
-\left(\frac{\partial\widetilde{\psi}}
{\partial x}\right)^2(x,\zeta,w)}{\left(\widetilde{\psi}
(x,\zeta,w)\right)^2}\right| 
& \geq & \frac{C_{\infty}(l,1)}{2((N+M+1)\log(K+K^2))^l}\\
& \geq & \frac{\widetilde{C}}{((N+M)\log(N+M))^l}\,.
\end{eqnarray*}

On the other case one has
$\left|\frac{\widetilde{\psi}(x,\zeta,w)}{\widetilde{\psi}
(0,\zeta,w)}\right|<e^{-\frac{b_{N,M}}{2}}$.
After moving $x$ closer to $0$ if necessary we can assume that the
function 
$t\mapsto\widetilde{\psi}(t,\zeta,w)$ 
does not vanish on $[0,x]$~: if it si not the case one can consider
$x_0$ as the first zero 
($>0$) of $\widetilde{\psi}(\cdot,\zeta,w)$. Since
$\lim_{t\rightarrow x_0^-}\widetilde{\psi}(t,\zeta,w)=0$, there is
$x$ sufficiently close to $x_0$, such that
$$0<\left|\frac{\widetilde{\psi}(x,\zeta,w)}{\widetilde{\psi}
(0,\zeta,w)}\right|<e^{-\frac{b_{N,M}}{2}}.$$
Moreover $\widetilde{\psi}(x,\zeta,w)$ and
$\widetilde{\psi}(0,\zeta,w)$ have same sign, the function
$$t\in[0,x]\mapsto\ln\frac{\widetilde{\psi}(t,\zeta,w)}{\widetilde{\psi}
(0,\zeta,w)},$$
is well defined (and of class $C^2$). Since
$$\ln\frac{\widetilde{\psi}(x,\zeta,w)}{\widetilde{\psi}(0,\zeta,w)}
<-\frac{b_{N,M}}{2},$$
there is $x_1\in[0,x]$, such that
$$\left|\left(\frac{\partial}{\partial t}
\ln\frac{\widetilde{\psi}(t,\zeta,w)}
{\widetilde{\psi}(0,\zeta,w)}\right)
(x_1,\zeta,w)\right|=\left|\frac{\frac{\partial\widetilde{\psi}}
{\partial t}(x_1,\zeta,w)}{\widetilde{\psi}(x_1,\zeta,w)}\right|>
\frac{b_{N,M}}{2}.$$
Otherwise we would have
$$\left|\ln\frac{\widetilde{\psi}(x,\zeta,w)}{\widetilde{\psi}
(0,\zeta,w)}\right|=\left|\int_0^x\frac{\frac{\partial\widetilde{\psi}}
{\partial t}(t,\zeta,w)}{\widetilde{\psi}(t,\zeta,w)}dt\right|
\leq\frac{b_{N,M}}{2},$$
which is impossible.

And since from hypothesis 
$\frac{\partial}{\partial t}\left(e^{\widetilde{k}(t,\zeta,w)}
\widetilde{\psi}(t,\zeta,w)\right)(0)=0$, there is
as well $x_2\in[0,x_1]$, such that
$$\left|\frac{\frac{\partial^2\widetilde{\psi}}{\partial^2t}(x_2,\zeta,w)
\widetilde{\psi}(x_2,\zeta,w)-\left(\frac{\partial\widetilde{\psi}}
{\partial t}\right)^2(x_2,\zeta,w)}{\left(\widetilde{\psi}(x_2,\zeta,w)
\right)^2}\right|>\frac{b_{N,M}}{2}\,,$$
thus
$$\left|h(x_2)-\frac{1}{b_{N,M}}\frac{\frac{\partial^2\widetilde{\psi}}
{\partial^2t}(x_2,\zeta,w)\widetilde{\psi}(x_2,\zeta,w)
-\left(\frac{\partial\widetilde{\psi}}{\partial t}\right)^2(x_2,\zeta,w)}
{\left(\widetilde{\psi}(x_2,\zeta,w)\right)^2}\right|\geq
\frac{1}{2}-\|h\|_{\infty}\geq\frac{1}{4},$$
since $2^lC_{\infty}(l,1)\leq\frac{1}{4}$ (see the statement of
theorem~\ref{vitushkin}). It follows that
$$\left|h(x_2)-\frac{1}{b_{N,M}}\frac{\frac{\partial^2\widetilde{\psi}}
{\partial^2t}(x_2,\zeta,w)\,\widetilde{\psi}(x_2,\zeta,w)-
\left(\frac{\partial\widetilde{\psi}}{\partial t}\right)^2(x_2,\zeta,w)}
{\left(\widetilde{\psi}(x_2,\zeta,w)\right)^2}\right|\geq\frac{1}{4}
\geq\frac{\widetilde{C}}{((N+M)\log(N+M))^l}$$
($\widetilde{C}$ is small enough), which completes the second case.
\bigskip

At last
$$\frac{\partial^2}{\partial x^2}\ln\frac{\widetilde{\psi}(x,\zeta,w)}
{\widetilde{\psi}(0,\zeta,w)}=k(x,\zeta,w)+\frac{\partial^2}{\partial x^2}
\ln\frac{\psi(x,\zeta,w)}{\psi(0,\zeta,w)}\,,$$
we can conclude that there is $h\in\Lambda_l\cap C^l([0,1])$
which is identically zero on $\left[0,\frac{1}{20}\right]$
such that for all $(\zeta,w)\in\Omega_{N,M}$,
$$\left\|h-\frac{1}{b_{N,M}}
\left(k(\cdot,\zeta,w)+\frac{\frac{\partial^2\psi}
{\partial^2x}(\cdot,\zeta,w)\psi(\cdot,\zeta,w)-\left(\frac{\partial\psi}
{\partial x}\right)^2(\cdot,\zeta,w)}{\left(\psi(\cdot,\zeta,w)\right)^2}
\right)\right\|_{\infty}\geq\frac{\widetilde{C}}{((N+M)\ln(N+M))^l}\,.$$

\end{proof}

In a more particular case there is the following corollary :

\begin{corollary}\label{inverse2}

Consider here $\psi(x,\zeta)$ on $[0,1]\times\mathbb{C}^N$, of class
$C^1$ with respect to $x$ and such that $\psi(x,\zeta)$ and 
$\frac{\partial\psi}{\partial x}(x,\zeta)$ satisfy the conditions
of theorem~\ref{th1bis} with $M=0$, and
$$\frac{1}{\psi(0,\zeta)}=O\left(e^{\alpha N^{\beta}}\right).$$

Then the approximation of the compact set
$$\int_0^{\cdot}\Lambda_l([0,1])=\left\{\left(x\mapsto\int_0^xh(t)dt\right),
\,\,h\in\Lambda_l\right\}$$
by the family
$$\left\{\left(x\in[0,1]\mapsto\frac{1}{b_N}\left(\frac{1}{\psi}
\frac{\partial\psi}{\partial x}\right)(x,\zeta)\right),\,
\zeta_j=O(N^r),\,\forall\,j=1,\ldots,N\right\},$$
uniformly on $[0,1]$, where $b_N>0$ and $\frac{1}{b_N}$ are of
polynomial kind at $N$, cannot be better than
$$\frac{\widetilde{C}}{N^{l+1}(\ln N)^{l+1}}.$$

Moreover one can choose $h\in\Lambda_l\cap C^l(I^s)$ identically
zero on $\left[0,\frac{1}{20}\right]$.

\end{corollary}

\begin{proof}

We can in the same way assume that
$\forall\,\zeta\in\mathbb{R}^N,\;\psi(0,\zeta)>0$. Then
$\psi$ being given, let consider
$$\chi(x,\zeta_1,\ldots,\zeta_N,\zeta_{N+1})=e^{\zeta_{N+1}}
\frac{\partial\psi}{\partial x}(x,\zeta_1,\ldots,\zeta_N).$$
By theorem~\ref{th1bis} with $M=0$ there is $h_1\in\Lambda_{l+1}\cap
C^{l+1}([0,1])$ identically zero on $\left[0,\frac{1}{20}\right]$
which satisfies $(\ast')$ with respect to $\chi$.

Let consider $\mu\in[0,1]$, there is
\begin{eqnarray*}
|h_1(x)-\mu\chi(x,\zeta,\zeta_{N+1})| & \geq & 
|h_1(x)-\mu P_K(x,\zeta)|-\mu\|R_K(\cdot,\zeta,\zeta_{N+1})\|_{\infty}\\
& \geq & \frac{C_{\infty}(l+1,1)}{2((N+1)\log K)^{l+1}}\\
& \geq & \frac{\widetilde{C}}{(N\log N)^{l+1}}\,.
\end{eqnarray*}

Particularly one can choose
$\zeta_{N+1}=\ln\frac{e^{b_N}}{b_N\psi(0,\zeta)}=O\left(N^{r'}\right)$.
Now if $e^{b_N}\frac{\psi(x,\zeta)}{\psi(0,\zeta)}\geq1$, we get by
substituting $\mu=\frac{\psi(0,\zeta)}{e^{b_N}\psi(x,\zeta)}$
$$\left|h_1(x)-\frac{1}{b_N}\frac{\frac{\partial\psi}{\partial x}(x,\zeta)}
{\psi(x,\zeta)}\right|\geq\frac{\widetilde{C}}{(N\log N)^{l+1}}\,;$$
otherwise
$\psi(x,\zeta)<\frac{\psi(0,\zeta)}{e^{b_N}}$. Thus after taking $x$
close enough to $0$ if necessary, we can assume that the function 
$t\mapsto\psi(t,\zeta)$ 
does not vanish on $[0,x]$ : if it is not the case let consider
$x_0$ to be the first zero ($>0$) of $\psi(\cdot,\zeta)$. Since
$\lim_{t\rightarrow x_0^-}\psi(t,\zeta)=0$, there is an
$x_1$ sufficiently close to $x_0$ such that
$$0<\left|\frac{\psi(x_1,\zeta)}{\psi(0,\zeta)}\right|<e^{-b_N}.$$
Moreover $\psi(x_1,\zeta)$ and
$\psi(0,\zeta)$ have same sign, and so the function
$$t\in[0,x_1]\mapsto\ln\frac{\psi(t,\zeta)}{\psi(0,\zeta)}$$
is well defined (and of class $C^1$).

Since
$$\ln\frac{\psi(x_1,\zeta)}{\psi(0,\zeta)}<-b_N,$$
there is $x_2\in[0,x_1]$ such that
$$\left|\left(\frac{\partial}{\partial t}\ln\frac{\psi(t,\zeta)}
{\psi(0,\zeta)}\right)(x_2,\zeta)\right|=\left|\frac{\frac{\partial\psi}
{\partial t}(x_2,\zeta)}{\psi(x_2,\zeta)}\right|>b_N\,.$$
Otherwise we would have
$$\left|\ln\frac{\psi(x_1,\zeta)}{\psi(0,\zeta)}\right|
=\left|\int_0^{x_1}\frac{\frac{\partial\psi}
{\partial t}(t,\zeta)}{\psi(t,\zeta)}dt\right|\leq b_N,$$
which is impossible. And since $\|h_1\|_{\infty}\leq\frac{1}{2}$, we get
$$\left|h_1(x_1)-\frac{1}{b_N}
\frac{\frac{\partial\psi}{\partial x}(x_1,\zeta)}
{\psi(x_1,\zeta)}\right|\geq\frac{1}{2}
\geq\frac{\widetilde{C}}{(N\log N)^{l+1}}\,.$$

At last, set $h=h_1'\in\Lambda_l\cap C^l([0,1])$, then
$h_1(x)=\int_0^xh(t)dt$ (since $h_1(0)=0$) and $h$ is identically zero
on $\left[0,\frac{1}{20}\right]$ too.

\end{proof}

In the section~\ref{application} we will also use the following

\begin{corollary}\label{detail}

The corollaries~\ref{inverse1} and~\ref{inverse2} are still true
(with $l>1$) if we only consider in $\Lambda_l\cap C^{[l]}([0,1])$ the
functions $h>0$ which are strictly decreasing and satisfy
$$h^{(j)}(0)=0,\,\forall\,j=1,\ldots,[l].$$

\end{corollary}

\begin{proof}

Consider $\psi$ and set
$$\widetilde{\psi}(x,\zeta,w)=\exp\left(b_{N,M}\left(ax^{[l]+3}-dx^2\right)
\right)\,\psi(x,\zeta,w),$$
with $a,\,d>0$. Then $\widetilde{\psi}$ still fulfills the conditions of
corollariy~\ref{inverse1} (resp. corollary~\ref{inverse2})~: indeed
$b_{N,M}$ (resp. $b_N$) is polynomial 
at $N+M$ (resp. $N$), $\widetilde{\psi}$ is of class $C^2$ on
$[0,1]$ (resp. $C^1$) and
$$\frac{\partial^2}{\partial x^2}\left(\ln\frac{\widetilde{\psi}(x,\zeta,w)}
{\widetilde{\psi}(0,\zeta,w)}\right)=\frac{\partial^2}{\partial x^2}
\left(\ln\frac{\psi(x,\zeta,w)}{\psi(0,\zeta,w)}\right)
+b_{N,M}\left(a([l]+3)([l]+2)x^{[l]+1}-2d\right)\,.$$
According to the corollary~\ref{inverse1} there is
$h\in\Lambda_l\cap C^{[l]}([0,1])$ which satisfies :
$\forall\,(\zeta,w)\in\Omega_{N,M}$, $\exists\,x\in[0,1]$ such that
\begin{eqnarray*}
\left\|h-\frac{1}{b_{N,M}}\left(k+\frac{\partial^2}{\partial x^2}\ln
\frac{\widetilde{\psi}}{\widetilde{\psi}(0,\zeta,w)}\right)(\cdot,\zeta,w)
\right\|_{\infty} & = & \left\|\widetilde{h}
-\frac{1}{b_{N,M}}
\left(k+\frac{\partial^2}{\partial x^2}
\ln\frac{\psi}{\psi(0,\zeta,w)}\right)
(\cdot,\zeta,w)\right\|_{\infty}\\
& \geq & \frac{\widetilde{C}}{(N\log N)^l}\,,
\end{eqnarray*}
with $\widetilde{h}(x)=h(x)+2d-a([l]+3)([l]+2)x^{[l]+1}$.

In the case of corollary~\ref{inverse2} we get in the same way :
$\forall\,\zeta$, $\exists\,x\in[0,1]$ such that
\begin{eqnarray*}
\left|\int_0^xh(t)dt-\frac{1}{b_N\widetilde{\psi}(x,\zeta)}
\frac{\partial\widetilde{\psi}}{\partial x}(x,\zeta)\right| & =
& \left|\int_0^x\widetilde{h}(t)dt
-\frac{1}{b_N\psi(x,\zeta)}
\frac{\partial\psi}{\partial x}(x,\zeta)\right|\\
& \geq & \frac{C'}{(N\log N)^{l+1}}\,.
\end{eqnarray*}

Now we have to choose suitably $a$ and $d$ : on
$\left[\frac{1}{20},1\right]$, 
\begin{eqnarray*}
\widetilde{h}'(x) & = & h'(x)-a([l]+3)([l]+2)([l]+1)x^{[l]}\\
& \leq & -a([l]+3)([l]+2)([l]+1)\frac{1}{20^{[l]}}+\|h'\|_{\infty}\\
& < & 0\,,
\end{eqnarray*}
with $a$ sufficiently large (since $h\in\Lambda_l([0,1])$). Thus
$\widetilde{h}$ is strictly decreasing, as well as on
$\left[0,\frac{1}{20}\right]$ where $h$ is a constant. $a$ being fixed,
$d$ can be chosen sufficiently large in order to get
$\widetilde{h}>0$. It follows that $\widetilde{h}$ is strictly decreasing
and all its derivatives until order $[l]$ vanish at $0$.

Because $h$ is in a homothetic of $\Lambda_l([0,1])$, it is sufficient
to finish the proof to consider
$\frac{\widetilde{h}}{a([l]+3)([l]+2)+2d+1}$, after changing $b_{N,M}$
(resp. $b_N$)
and reducing $\widetilde{C}$ with respect to $a$ and $d$ (which only
depend on $[l]$).

\end{proof}

\section{An estimation of the eigenvalues and characteristic
  values}\label{estimation}

We consider the equation on the half-axis $\mathbb{R}^+$
$$-y''(x)-\omega^2Q(x)=\lambda y(x),$$
with the following hypothesis : $Q$ is strictly positive, integrable
and has $m+1$ derivatives which are polynomially decreasing at
infinity, and $\omega$ is a big parameter. 

For any sufficiently large $\omega$ the Sturm-Liouville operator
$-\frac{d^2}{dx^2}-\omega^2Q$ has $N(\omega)$ discrete strictly
negative eigenvalues $\lambda_j=-\xi_j^2$, $0<\xi_1<\ldots<\xi_N$, and
$N(\omega)$ eigenfunctions 
$\varphi_j$ which satisfy the condition :
$$\varphi_j(0)=0\text{ and }\int_{0}^{\infty}|\varphi_j(x)|^2dx=1.$$
Moreover the number $N(\omega)$ has the same order as
$\omega$ ; more precisely there are the bounds of Calogero
(cf.~\cite{calogero}) :
$$\frac{\omega}{\pi\sqrt{Q(0)}}\int_0^{\infty}Q(x)dx-\frac{1}{2}
\leq N(\omega)\leq\frac{2\omega}{\pi}\int_0^{\infty}\sqrt{Q(x)}dx\,.$$
It follows that for all sufficiently large
$\omega$, $a\omega\leq N(\omega)\leq b\omega$.

At last we set $C_j=\left(\varphi'_j(0)\right)^2$, the
characteristic value associated to $\xi_j$.
\bigskip

To apply at section~\ref{application} the negative results from
section~\ref{rappel}, we have to prove the following estimations :

\begin{proposition}\label{proofestim}

Let $q=-\omega^2Q$ be a strictly negative potential where $Q>0$ is
integrable of class 
$C^1$ with $Q'(0)=0$, strictly decreasing on $\mathbb{R}^+$ with
polynomial behavior at infinity. Then for all sufficiently large
$\omega$ and $j=1,\ldots,N(\omega)$, there are
$$\frac{a}{\omega^b}\leq\xi_j\leq c\omega\text{ and }
\frac{1}{\alpha\exp\left(\beta\omega^2\right)}\leq
\frac{4\xi_j^2}{C_j}\leq\alpha \exp\left(\beta\omega^{\gamma}\right),$$
where  $a,b,c,\alpha,\beta,\gamma$ are constants wich only depend on
$Q$.

\end{proposition}

\begin{proof}

This proposition is a corollary of the WKB theory which is not
precisely formulated in the references, so we give a
proof by using principally the WKB method like
in~\cite{lax-levermore}. Its consists in some following lemmas.

\begin{lemma}\label{eigenvalue}

For all $j=1,\ldots,N(\omega)$, 
$$\frac{a}{\omega^b}\leq\xi_j\leq c\omega\,.$$

\end{lemma}

\begin{proof}

Let $\phi_j$ be the normed eigenfunction then :
\begin{eqnarray*}
-\xi_j^2\,=\,\lambda_j & = & -\int_0^{\infty}\phi_j''(x)\phi_j(x)dx
-\omega^2\int_0^{\infty}Q(x)\,\phi_j^2(x)dx\\
& \geq &
-\left[\phi_j'\phi_j\right]_0^{\infty}+\int_0^{\infty}\phi_j'^2(x)dx
-\omega^2\inf_{x\geq0}Q(x)\\
& \geq & -\omega^2\sup_{x\geq0}Q(x),
\end{eqnarray*}
then $\xi_j\leq\omega\sup_{x\geq0}\sqrt{Q(x)}=\omega\sqrt{Q(0)}$.
\bigskip

About the lower estimate, we have to extend the equation on the whole
line $\mathbb{R}$ in order to apply the WKB method of Lax and
Levermore (see~\cite{lax-levermore}) : first we extend $Q$ on an even
function which still is of class $C^1$ (since
$Q'(0)=0$) and integrable with polynomial decreasing at
infinity. Moreover the extension (which we still write $Q$) is
monotone on $\mathbb{R}^-,\,\mathbb{R}^+$ and has the one maximum at
$0$.

Next we extend each eigenfunction $\phi_j$ on an odd function (this is
possible since $\phi_j(0)=0$) which still is of class $C^2$ on
$\mathbb{R}$ because $\phi_j''(0)=0$. So each extension is an
eigenfunction with the same eigenvalue, and conversely we see that by
restriction we get all of them (by uniqueness since the $\xi_j$ are
eigenvalues).

Now we consider the equation in the quasi-classic case :
$$-\varepsilon^2y''-Qy=-\eta^2y,$$
where $\varepsilon=\frac{1}{\omega}$ and
$\eta_j=\frac{\xi_{N-j+1}}{\omega}$, $j=1,\ldots,N$. Then
$$0\leq\eta_N\leq\ldots\leq\eta_1\leq\sqrt{Q(0)},$$
and by the WKB method
$$\Phi(\eta_j)=\left(j-\frac{1}{2}\right)\varepsilon\pi,$$
where
$$\Phi(\eta)=\int_{x_-(\eta)}^{x_+(\eta)}\left(Q(y)
-\eta^2\right)^{\frac{1}{2}}dy$$
with $x_+(\eta)=-x_-(\eta)\geq0$ satisfying
$-Q(x_+(\eta))=-Q(x_-(\eta))=-\eta^2$ (well-defined by monotonicity of
$Q$ on $\mathbb{R}^-,\;\mathbb{R}^+$). Next
$$N(\omega)=N(\varepsilon)=\left[\frac{1}{\varepsilon\pi}\Phi(0)\right]
=\left[\frac{1}{\varepsilon\pi}\int_{-\infty}^{\infty}
(Q(y))^{\frac{1}{2}}dy\right].$$
At last
$$s_j=s(\eta_j)=\exp\left(\frac{\theta_+(\eta_j)}{\varepsilon}\right)$$
where
$$\theta_+(\eta)=\eta x_+(\eta)+\int_{x_+(\eta)}^{\infty}\eta
-\left(\eta^2-Q(y)\right)^{\frac{1}{2}}dy$$
and $s_j$ is the normalized coefficient of the Jost solution :
$$\phi_j(x)\sim s_j\exp\left(-\frac{\eta_j x}{\varepsilon}\right).$$
\bigskip

Hence it is sufficient to prove that
$$\eta_N\geq\frac{a}{\omega^b}$$
(for more simplicity we call by $a$ different constants). There is from
hypothesis for $x$ large enough
$$\frac{1}{a}\frac{1}{|x|^{k_1}}\leq Q(x)\leq\frac{a}{|x|^{k_2}},$$
(with $k_1\geq k_2\geq4$) and since
$\eta_{N}\rightarrow0$ when $\omega\rightarrow\infty$, then 
$$\frac{1}{a}\frac{1}{x_+(\eta_N)^{k_1}}\leq
Q(\eta_N)=\eta_N^2\leq\frac{a}{(x_+(\eta_N))^{k_2}},$$
and
$$\frac{1}{a}\frac{1}{{\eta_N}^{\frac{2}{k_1}}}\leq
x_+(\eta_N)\leq\frac{a}{{\eta_N}^{\frac{2}{k_2}}}$$
(as well as $x_-(\eta_N)=-x_+(\eta_N)$ ). Thus
$$\Phi(0)-\Phi(\eta_N)=\left(\int_{-\infty}^{x_-(\eta_N)}
+\int_{x_+(\eta_N)}^{\infty}\right)(Q(y))^{\frac{1}{2}}dy\;
+\;\int_{x_-(\eta_N)}^{x_+(\eta_N)}(Q(y))^{\frac{1}{2}}
-\left(Q(y)-\eta_N^2\right)^{\frac{1}{2}}dy.$$
First one has
\begin{eqnarray*}
\int_{x_+(\eta_N)}^{\infty}(Q(y))^{\frac{1}{2}}dy
& \leq & \int_{x_+(\eta_N)}^{\infty}\frac{a}{y^{\frac{k_2}{2}}}dy
=\frac{a}{x_+(\eta_N)^{\frac{k_2-2}{2}}},\\
& \leq & a\,{\eta_N}^{\frac{k_2-2}{k_1}}
\end{eqnarray*}
as well as
$\int_{-\infty}^{x_-(\eta_N)}(Q(y))^{\frac{1}{2}}dy$.
Next
\begin{eqnarray*}
\int_{x_-(\eta_N)}^{x_+(\eta_N)}(Q(y))^{\frac{1}{2}}
-\left(Q(y)-\eta_N^2\right)^{\frac{1}{2}}dy
& \leq & \int_{x_-(\eta_N)}^{x_+(\eta_N)}\left(Q(y)-\left(Q(y)
-\eta_N^2\right)\right)^{\frac{1}{2}}dy\\
& = & \eta_N(x_+(\eta_N)-x_-(\eta_N))=a\,{\eta_N}^{1-\frac{2}{k_2}}\\
& \leq & a\,{\eta_N}^{\frac{k_2-2}{k_1}}
\end{eqnarray*}
(because $k_1\geq k_2$ and $\eta_N<1$ for $\omega$ large enough), thus
$$\Phi(0)-\Phi(\eta_N)\leq a\,{\eta_N}^{\frac{k_2-2}{k_1}}\,.$$
Since
$N=\left[\frac{1}{\varepsilon\pi}\Phi(0)\right]\leq
\frac{\Phi(0)}{\varepsilon\pi}$
there is
$$\Phi(0)-\Phi(\eta_N)\geq\frac{\varepsilon\pi}{2},$$
it follows that
$$a{\eta_N}^{\frac{k_2-2}{k_1}}\geq\frac{\varepsilon\pi}{2},$$
then
$$\eta_N\geq a\varepsilon^{\frac{k_1}{k_2-2}}
=\frac{a}{\omega^{\frac{k_1}{k_2-2}}}\,,$$
which completes the proof of the lemma.

\end{proof}

In the following we need a formula of $C_j$ (see~\cite{chadan}). We
extend the equation
$$-y''-\omega^2Qy=k^2y$$
with respect to $k$ in the half-plane $\left\{\Im m\,k>0\right\}$. We
know that there are the solutions $\phi(k,\cdot),\,\psi(k,\cdot)$ and
$f(k,\cdot)$ which respectively fulfill
$$
\begin{cases}
\phi(k,\cdot)\in L^2([0,\infty[),\text{ and
}\int_0^{\infty}|\phi(k,x)|^2dx=1\\
\psi(k,0)=0\;\text{ and }\;\psi'(k,0)=1\\
f(k,x)\sim\exp(ikx),\;x\rightarrow\infty\,.\\
\end{cases}
$$
$\phi(k,\cdot)$ is the physical solution and $f(k,\cdot)$ is called
the Jost solution. When $k$ is one of the eigenvalues $\xi_j$ (and
only in this case) these three solutions are proportional (and can be
chosen to be real) and there is
$$
\begin{cases}
f(k_j,\cdot)=f'(k_j,0)\psi(k_j,\cdot),\\
\phi(k_j,\cdot)=C_j^{\frac{1}{2}}\psi(k_j,\cdot),\\
\phi(k_j,\cdot)=s_jf(k_j,\cdot)
\end{cases}
$$
(the solutions $\phi(k,\cdot)$ and $f(k,\cdot)$ are always
proportional since $\Im m\,k>0$), then
$$f'(k_j,0)=\frac{C_j^{\frac{1}{2}}}{s_j}\,.$$
In addition we define the Jost function $F$ as $F(k)=f(k,0)$. From
above we know that its zeros are exactly the eigenvalues (as we will
see $F$ is holomorphic on the half-plane and its zeros are
simple). Now we claim that
$$\frac{4\xi_j^2}{C_j}=-s_j^2\left(\dot{F}(i\xi_j)\right)^2.$$
Indeed by differentiating with respect to $k$ the equation
$$-f''(k,x)-\omega^2Q(x)f(k,x)=k^2f(k,x),$$
the derivative (with respect to $x$) of the wronskian of $f(k,\cdot)$
and $\dot{f}(k,\cdot)=\frac{\partial f}{\partial k}(k,\cdot)$ is
\begin{eqnarray*}
W'\left(f,\dot{f}\right) & = &
\left(f(k,x)\dot{f}'(k,x)-f'(k,x)\dot{f}(k,x)\right)'\\
& = & f(k,x)\dot{f}''(k,x)-f''(k,x)\dot{f}(k,x)\\
& = & -2kf^2(k,x),
\end{eqnarray*}
which gives when $k=k_j$
\begin{eqnarray*}
\dot{F}(k_j)f'(k_j,0) & = &
-2k_j\int_0^{\infty}f^2(k_j,x)dx=-\frac{2k_j}{s_j^2},\\
& = & \dot{F}(k_j)\frac{C_j^{\frac{1}{2}}}{s_j},
\end{eqnarray*}
this proves the assertion.
\bigskip

We see in addition (since $0<\eta_N<\ldots<\eta_1\leq\sqrt{Q(0)}$ )
that $0\leq x_+(\eta_1)\leq\ldots\leq x_+(\eta_N)$ and for all
$j=1,\ldots,N(\omega)$,
$$\theta_+(\eta_j)\leq x_+(\eta_N)
+\int_0^{\infty}(Q(y))^{\frac{1}{2}}dy$$
which gives from lemma~\ref{eigenvalue}
$$\theta_+(\eta_j)\leq a\,\omega^{\frac{2k_1}{k_2(k_2-2)}},$$
then
$$s_j\leq\exp\left(a\omega^{1+\frac{2k_1}{k_2(k_2-2)}}\right)\,.$$
Since $s_j\geq1$ it is enough to estimate
$\left|\dot{F}(i\xi_j)\right|$ hence the proof of the proposition is
complete thanks to the following lemma.

\end{proof}

\begin{lemma}

For all $j=1,\ldots,N(\omega)$
$$\frac{1}{a\exp(b\omega^2)}\leq\left|\dot{F}(i\xi_j)\right|\leq
a\exp\left(b\omega^2\right).$$

\end{lemma}

\begin{proof}

First let consider the equation
$$y''-Vy=-k^2y$$
where $x\geq0$, $V(x)=-\omega^2Q(x)$ and $\Im m\,k\geq0$.
The Jost solution can be constructed by successive approximations by setting
$$
\begin{cases}
f_0(k,x)=e^{ikx},\\
f_n(k,x)=\int_x^{\infty}\frac{\sin k(x-t)}{k}V(t)f_{n-1}(t)dt\,,\;n\geq1,
\end{cases}
$$
which gives
$$f(k,\cdot)=\sum_{n=0}^{\infty}f_n(k,\cdot),$$
and $F(k)=f(k,0)$ for the Jost function. We know (see~\cite{chadan})
that $F$ is
holomorphic on the half-plane $\{\Im m\,k>0\}$, continue on 
$\{\Im m\,k\geq0\}$, vanishes exactly on the eigenvalues $i\xi_j$ and
converges to $1$ at infinity. Moreover there is the following estimate
: $\forall\,x,\,k$,
$$|f_n(k,x)|\leq\exp(-\Im m\,kx)\,\frac{1}{n!}
\left(\int_x^{\infty}\frac{2\sqrt{2}\,t}{1+|k|t}|V(t)|dt\right)^n,$$
then for all $k$
\begin{eqnarray*}
|F(k)|\leq\sup_{x\in\mathbb{R}}|f(k,x)| & \leq & 
\exp\left(2\sqrt{2}\,\omega^2\int_0^{\infty}t|Q(t)|dt\right)\\
& = & \exp\left(a\omega^2\right).
\end{eqnarray*}
For all $i\xi_j$ one has by the Cauchy formula on a small disc
$D(i\xi_j,\varepsilon)$ which is contained on the domain of holomorphy
of $F$,
$$\left|\dot{F}(i\xi_j)\right|=\left|\frac{1}{2\pi i}
\int_{|k-i\xi_j|=\varepsilon}\frac{F(k)}{\left(k-i\xi_j\right)^2}dk\right|
\leq\frac{\|F\|_{\infty}}{\varepsilon}.$$
One can choose $\varepsilon=\frac{\xi_1}{2}$ with $\xi_1=\omega\eta_N$
being the smallest eigenvalue (which is $\geq\frac{b}{\omega}$ by
lemma~\ref{eigenvalue}), then for all $j=1,\ldots,N(\omega)$
$$\left|\dot{F}(i\xi_j)\right|\leq b\omega^c\exp\left(a\omega^2\right)\leq
b'\exp\left(a'\omega^2\right)$$
and this proves the upper estimate.
\bigskip

For the lower estimate we set
$$\widetilde{F}(k)=\frac{F(k)}{\prod_{l=1}^N\frac{k-i\xi_l}{k+i\xi_l}}\,,$$
which is as well continue on the closed half-plane, holomorphic inside,
converges to $1$ at infinity and does not vanish. Moreover
\begin{eqnarray*}
\dot{F}(i\xi_j) & = &
\frac{d}{dk}\left(\prod_{l=1}^N\frac{k-i\xi_l}{k+i\xi_l}\right)(i\xi_j)\;
\widetilde{F}(i\xi_j)\\
& = & \frac{1}{2i\xi_j}\left(\prod_{l\neq j}
\frac{\xi_j-\xi_l}{\xi_j+\xi_l}\right)\widetilde{F}(i\xi_j).
\end{eqnarray*}
First we know that $\xi_j\leq\xi_N\leq b\omega$. Next in order to get
a lower estimate of $|\xi_j-\xi_l|$ (we could think that the
distribution of the eigenvalues is asymptotically uniform), it is
enough to show that 
$$\forall\,j,\,l\text{ with }1\leq j<l\leq N,\,\eta_j-\eta_l
\geq\frac{a}{\omega^b}\,.$$
Indeed one has
$$\eta_j-\eta_l\geq\min_{1\leq l\leq N-1}(\eta_l-\eta_{l+1})$$
and
$$\frac{\pi}{\omega}=\Phi(\eta_{l+1})-\Phi(\eta_l)
=\int_{x_-(\eta_{l+1})}^{x_+(\eta_{l+1})}\left(Q(y)
-\eta_{l+1}^2\right)^{\frac{1}{2}}dy-\int_{x_-(\eta_l)}^{x_+(\eta_l)}
\left(Q(y)-\eta_l^2\right)^{\frac{1}{2}}dy.$$
On $[x_+(\eta_l),x_+(\eta_{l+1})]$
(resp. $[x_-(\eta_{l+1}),x_-(\eta_l)]$) there is
$\eta_{l+1}^2\leq Q(y)\leq\eta_l^2$ then
$$Q(y)-\eta_{l+1}^2\leq\eta_l^2-\eta_{l+1}^2,$$
and on $[x_-(\eta_l),x_+(\eta_l)]$
$$\left(Q(y)-\eta_{l+1}^2\right)^{\frac{1}{2}}-\left(Q(y)
-\eta_l^2\right)^{\frac{1}{2}}\leq\left(\eta_l^2
-\eta_{l+1}^2\right)^{\frac{1}{2}},$$
therefore
\begin{eqnarray*}
\frac{\pi}{\omega} & \leq & (\eta_l-\eta_{l+1})^{\frac{1}{2}}
(\eta_l+\eta_{l+1})^{\frac{1}{2}}(x_+(\eta_{l+1})-x_-(\eta_{l+1}))\\
& \leq & \sqrt{2}\,(\eta_l-\eta_{l+1})^{\frac{1}{2}}(x_+(\eta_{l+1})
-x_-(\eta_{l+1})).
\end{eqnarray*}
On the other side 
$$x_+(\eta_{l+1})-x_-(\eta_{l+1})\leq x_+(\eta_N)
-x_-(\eta_N)\leq a\,\omega^b$$
then 
$$\eta_l-\eta_{l+1}\geq\frac{a}{\omega^{b'}}\,.$$
It follows that (since $N(\omega)=O(\omega)$ )
$$\prod_{l\neq j}\left|\frac{\xi_j-\xi_l}{\xi_j+\xi_l}\right|
\geq\left(\frac{a}{\omega^{b'}}\right)^{N-1}
\geq\frac{a''}{\exp(b''\omega^2)}\,.$$
\bigskip

Now we have to get a lower estimate for
$\left|\widetilde{F}(i\xi_j)\right|$. Since $\widetilde{F}$ is
holomorphic without zero one has for any $R$ large enough :
$$\frac{1}{4i\xi_j\,\widetilde{F}(i\xi_j)}
=\frac{1}{2\pi i}\left(\int_{-R}^R\frac{k}{\widetilde{F}(k)
(k-i\xi_j)(k+i\xi_j)^2}dk+\int_0^{\pi}
\frac{iR^2e^{2i\theta}}{\widetilde{F}(Re^{i\theta})
(Re^{i\theta}-i\xi_j)(Re^{i\theta}+i\xi_j)^2}d\theta\right)\,,$$
which gives by taking limit 
($\widetilde{F}(k)\rightarrow1,\,k\rightarrow\infty$)
$$\frac{1}{\widetilde{F}(i\xi_j)}
=\frac{4i\xi_j}{2\pi i}\int_{-\infty}^{+\infty}
\frac{k}{\widetilde{F}(k)(k-i\xi_j)(k+i\xi_j)^2}dk$$
thus
$$\frac{1}{\left|\widetilde{F}(i\xi_j)\right|}\leq
b\omega\int_0^{\infty}\frac{k}{|F(k)|\,|k-i\xi_j|\,
|k+i\xi_j|^2}dk,$$
since $\forall\,k\in\mathbb{R},\;\left|\widetilde{F}(k)\right|=|F(k)|$
and $F(-k)=\overline{F(k)}$.

In order to get a lower estimate of $\frac{1}{|F(k)|}$ on $\mathbb{R}^+$
we have to use the wronskian of
$f(k,\cdot)$ and $f(-k,\cdot)$ which is constant and equal to $-2ik$
(it can be calculated by taking the limit at infinity since
$f'(k,x)\sim ike^{ikx}$). In addition for all $k>0$, it is equal to
$$f(k,0)f'(-k,0)-f(-k,0)f'(k,0)
=2i\Im m\left(F(k)\overline{f'(-k,0)}\right).$$
And by the integral equation of which $f(k,\cdot)$ is solution,
$$f(k,x)=e^{ikx}+\int_x^{\infty}\frac{\sin k(x-t)}{k}V(t)f(k,t)dt,$$
we can deduce by the estimation of the Jost solution from above that
$\forall\,k>0$
\begin{eqnarray*}
|f'(k,0)| & \leq & 
k+\int_0^{\infty}|\cos kt|\,|V(t)|\,|f(k,t)|dt\\ 
& \leq & C\omega^2+\omega^2\exp\left(a\omega^2\right)
\int_0^{\infty}|Q(t)|dt\\
& \leq & C'\exp\left(a'\omega^2\right),
\end{eqnarray*}
as well for $f'(-k,0)=\overline{f'(k,0)}$. Thus
$$2k\leq2|F(k)|C'\exp\left(a'\omega^2\right)$$
which gives
\begin{eqnarray*}
\int_0^{+\infty}\frac{k}{|F(k)|\,|k-i\xi_j|\,|k+i\xi_j|^2}dk & = & 
C'\exp\left(a'\omega^2\right)\int_0^{+\infty}
\frac{1}{\left(k^2+\xi_j^2\right)^{\frac{3}{2}}}dk\\
& \leq & C'\exp\left(a'\omega^2\right)\left(\frac{1}{\xi_j^3}
+\int_1^{+\infty}\frac{1}{k^3}dk\right)\\
& \leq
& C''\omega^b\exp\left(a'\omega^2\right)\,.
\end{eqnarray*}
At last
$$\frac{1}{\left|\widetilde{F}(i\xi_j)\right|}\leq
C''\omega^{b'}\exp\left(a''\omega^2\right)\leq
\alpha\exp\left(\beta\omega^2\right)\,$$
and the proof is complete.

\end{proof}

\begin{remark}\label{constcomp}

The constants $a,b,c,\alpha,\beta,\gamma$ which appear in
the statement of the proposition only depend on any compact
$\Lambda_{\mathcal{Q}}\ni Q$ : indeed their depend on
$Q(0),\;\int_0^{\infty}(1+t)\sqrt{Q}(t)dt$ and $a,\;k_1,\;k_2$ (to
assume that $\frac{1}{a}\frac{1}{x^{k_1}}\leq
Q(x)\leq\frac{a}{x^{k_2}}$ ). This will be usefull to
section~\ref{application}.

In addition when we assume that $k_1=k_2=k$ there is for all
$j=1,\ldots,N(\omega)$, 
$$b\,\omega\geq\xi_j\geq\omega\frac{a}{\omega^{\frac{k}{k-2}}}
\geq\frac{a}{\omega}$$
and
$$\frac{1}{\alpha\,\exp(\beta\omega^2)}\leq\frac{\xi_j^2}{C_j}\leq\alpha\,
\exp\left(\beta\omega^2\right)\,.$$

\end{remark}

Before ending let consider two different examples.

\begin{example}

Let be
$$Q_1(x)=\frac{1}{\left(1+x^2\right)^2}.$$
$Q_1$ is even and satisfies the conditions of
proposition~\ref{proofestim}. By WKB method there is (with
$\omega\geq10$) :
$$N(\omega)=\left[\frac{1}{\varepsilon}\right]=[\omega]\,.$$
There is $x_+(\eta)=-x_-(\eta)=\sqrt{\frac{1}{\eta}-1}$ and
$$\frac{\pi^2}{256}\,\frac{1}{\omega^2}\leq\eta_N\leq\frac{\pi^2}{16}
\frac{1}{\omega^2},$$
then $\forall\,n$ with $1\leq n\leq N$, 
$$\frac{\pi^2}{256}\frac{1}{\omega}\leq\xi_n\leq\omega.$$
About the distribution of the eigenvalues one has
$$\xi_n-\xi_{n-1}\geq\frac{1}{5\omega}.$$
At last $\forall\,n$ with $1\leq n\leq N$,
$$1\leq s_n\leq\exp\left(\frac{4}{\pi}\omega^2+\frac{\pi}{2}\omega\right)$$
and
$$\frac{1}{\exp\left(2\omega\ln4\omega\right)}
\leq\prod_{j\neq n}\left(\frac{\xi_n-\xi_j}{\xi_n+\xi_j}\right)^2\leq1\,;$$
next
$$\left|\dot{F}(i\xi_j)\right|\leq\frac{512}{\pi^2}\omega
\exp\left(\pi\sqrt{2}\,\omega^2\right)\;\text{ and }\;
\frac{1}{\left|\widetilde{F}(i\xi_j)\right|}\leq2^{22}\omega^7
\exp\left(\pi\sqrt{2}\,\omega^2\right),$$
thus
$$\frac{1}{45\,\omega^9\exp\left(26\,\omega^2\right)}\leq
\frac{4\xi_n^2}{C_n}
\leq22\,\omega^2\exp\left(15\,\omega^2\right)\,.$$

\end{example}
\bigskip

\begin{example}

Let be
$$Q_2(x)=
\begin{cases}
1,\text{ si } x\in[0,1],\\
0,\text{ si } x>1.
\end{cases}
$$
The associate potential is a special case because it is discontinuous
at $x=1$. Then we calculate directly the estimations.

The equation $-y''-\omega^2y=\lambda y$ with $\lambda=-\xi^2$ and
$0<\xi<\omega$, has the following eigenfunctions
$$
y_{\xi}(x)=
\begin{cases}
\sqrt{\frac{2\xi}{1+\xi}}\sin\left(\sqrt{\omega^2-\xi^2}\,x\right),
\text{ if }x\in[0,1],\\
\sqrt{\frac{2\xi}{1+\xi}}\,e^{\xi}\sin\sqrt{\omega^2-\xi^2}\,
e^{-\xi x},\text{ if }x\geq1,
\end{cases}
$$
where $\xi$ satisfies the equation
$$\xi\sin\sqrt{\omega^2-\xi^2}+\sqrt{\omega^2-\xi^2}\cos
\sqrt{\omega^2-\xi^2}=0.\pod\diamond$$
The solutions $y_{\xi}\in C^1(\mathbb{R}^+)$ fulfill the conditions at
$0$ and $+\infty$, and are normed :
$$\int_0^{+\infty}y_{\xi}^2(x)dx=1.$$

First we know that $\xi=O(\omega)$. And since
$$C_{\xi}=\left(y_{\xi}'(0)\right)^2=\frac{2\xi}{1+\xi}
\left(\omega^2-\xi^2\right),$$
there is
$$\frac{4\xi^2}{C_{\xi}}=\frac{2\xi(\xi+1)}{\omega^2-\xi^2}.$$
About the first estimation we find as well
$0<\frac{4\xi^2}{C_{\xi}}\leq220\,\omega^2$ : indeed, if
$\omega\geq10$ and $\varepsilon=\sqrt{\omega^2-\xi^2}\leq\frac{1}{10}$,
the equation$\pod\diamond$ becomes
$$\sqrt{\omega^2-\varepsilon^2}\sin\varepsilon+\varepsilon\cos\varepsilon
\geq10\,\varepsilon>0.$$
It follows that $\varepsilon$ must be $\geq\frac{1}{10}$ then
$$\xi\leq\sqrt{\frac{99}{100}}\,\omega.$$

However the lower estimate of $\frac{4\xi^2}{C_{\xi}}$ need an
analogous estimate for $\xi$ which is in general false since the
first eigenvalue $\xi_1$ cannot be low bounded by
$\frac{1}{\omega^k}$ or $\exp(-a\omega^b)$, necessary condition to get
$\ln\frac{4\xi^2}{C_{\xi}}=O(\omega^b)$.

Indeed let fix for example $\omega_0$ large enough and
$\equiv\frac{\pi}{2}\pmod{2\pi}$, and consider the
equation$\pod\diamond$ with respect to $(\xi,\omega)$, given
by a smooth function $g$ in a neighbourhood of $(0,\omega_0)$. Since
$\frac{\partial g}{\partial\xi}(0,\omega_0)=\sin\omega_0=1$ and
$\frac{\partial g}{\partial\omega}(0,\omega_0)=-\omega_0$, an
application of the implicit functions theorem allows to
consider the function $\omega\mapsto\xi(\omega)$ which has in a
neighbourhood of $\omega_0$ the following expansion
$$\xi(\omega)=\omega_0(\omega-\omega_0)+O\left((\omega-\omega_0)^2\right).$$
Then we see that for all $\omega$ such that
$0<\omega-\omega_0\leq\eta(\omega_0)$ with $\eta(\omega_0)$
small enough, one has
$$0<\xi(\omega)\leq2\exp\left(-\frac{1}{2}\exp\omega\right)$$
and it follows that for any $\omega$ large enough we cannot find a lower
estimate for $\xi$ as we need. This accident arises from the fact that
$Q_2$ is not continuous at $x=1$.
\bigskip

However if we set a condition for $\omega\geq10$ like
$$\left|\omega-\frac{\pi}{2}+\pi\mathbb{Z}\right|\geq\frac{1}{5},$$
we see that for any $\xi$ with $0<\xi\leq\frac{1}{10}$, we get
$\left|\sqrt{\omega^2-\xi^2}-\frac{\pi}{2}+\pi\mathbb{Z}\right|
\geq\frac{1}{10}$ then
$$\left|\xi\sin\sqrt{\omega^2-\xi^2}+\sqrt{\omega^2-\xi^2}\cos
\sqrt{\omega^2-\xi^2}\right|\geq\frac{1}{2}.$$
Hence $\xi\geq\frac{1}{10}$ and
$$\frac{4\xi^2}{C_{\xi}}\geq\frac{1}{5\omega^2},$$
then we get as well a stronger estimation for $\frac{4\xi^2}{C_{\xi}}$
as $O(\omega^2)$.

\end{example}
\bigskip

\section{Some properties of the solution of a certain integral
  equation}\label{solequint}

Consider for $x,\,y\in\mathbb{R}^+$ and $w\in W=\left\{\Re e\,z>0\right\}$
$$\Phi(x,y,w)=\int_0^{\infty}\frac{\sin{kx}}{k}\frac{\sin{ky}}{k}
\left(\sqrt{k^2+w}-k\right)\frac{2k}{\pi}\,dk$$
and the integral equation with $(x,y)\in\Delta=\{0\leq y\leq x\}$ and
$w\in W$ :
\begin{equation}\label{equint}
A(x,y,w)+\int_0^xA(x,s,w)\Phi(s,y,w)ds+\Phi(x,y,w)=0\,.
\end{equation}
We want to prove in this section the following result :

\begin{proposition}\label{solution}

The equation~\ref{equint} has an unique solution $A(x,y,w)$ defined on
$\Delta\times W$. It is continuous with respect to $(x,y)$ 
and holomorphic with respect to $w$, and the application
$$x\mapsto\left(y\mapsto A(x,y,w)\in L^2_y([0,x])\right),$$
is continuously differentiable (with respect to the topology of
$L^2([0,x])$ ).

Moreover there are the following estimations : for all
$X\geq1$ and $w\in W$,
$$\sup_{y\leq x\leq X}|A(x,y,w)|,\;
\sup_{x\leq X}\left\|\frac{\partial A}{\partial x}(x,y,w)
\right\|_{L^2_y([0,x])},\;
\sup_{x\leq X}\left|\frac{d}{dx}A(x,x,w)\right|
\leq C(X)(1+|w|)^{\alpha}\,.$$
(the exponent $\alpha$ does not depend on $X$)

\end{proposition}

The proof consists on the following lemmas. We begin by proving these
properties about $\Phi$ :

\begin{lemma}\label{phi}

The function $\Phi$ is continuous with respect to
$(x,y)\in\mathbb{R}^+\times\mathbb{R}^+$ and holomorphic with respect to
$w\in W$. For all $X\geq0$ the application 
$$x\mapsto\Phi(x,y,w)\in L^2_y([0,X])$$
is continuously differentiable and there are the following estimates
:
$$\sup_{x,y\in[0,X]}|\Phi(x,y,w)|,\;
\sup_{x\leq X}\left\|\frac{\partial\Phi}{\partial x}
(x,y,w)\right\|_{L^2_y([0,X])}\leq C(X)(1+|w|)^{\alpha}.$$

Moreover the restriction
$$x\mapsto\Phi(x,x,w)$$
is continuously differentiable (in the usual sense) and there is the
following estimate~:
$$\sup_{x\leq X}\left|\frac{d}{dx}\Phi(x,x,w)\right|\leq 
C(X)(1+|w|)^{\alpha}.$$
(in all the cases the exponent $\alpha$ does not depend on $X$)

\end{lemma}

\begin{proof}

First $\Phi$ is well-defined since the integral is absolutely
convergent :
$$\Phi(x,y,w)=\int_0^{\infty}\frac{\sin{kx}}{k}\frac{\sin{ky}}{k}
\frac{w}{k+\sqrt{k^2+w}}\frac{2k}{\pi}dk,$$
where we have chosen the principal determination of $\sqrt{z}$. 
$\Phi$ is holomorphic with respect to $w$ and for all $X\geq1$ 
$$\|\Phi(\cdot,w)\|_{\infty,X}\leq C(X)|w|.$$
\bigskip

Next, we have to prove the differentiability of $x\mapsto\Phi(x,\cdot,w)$ :
$$\Phi(x,y,w)=\left(\int_0^1+\int_1^{\infty}\right)
\frac{\sin{kx}}{k}\frac{\sin{ky}}{k}
\frac{w}{k+\sqrt{k^2+w}}\frac{2k}{\pi}dk,$$
and
$$\frac{\partial}{\partial x}\int_0^1\frac{\sin{kx}}{k}\frac{\sin{ky}}{k}
\left(\sqrt{k^2+w}-k\right)\frac{2k}{\pi}dk
=\int_0^1\cos{kx}\frac{\sin{ky}}{k}\left(\sqrt{k^2+w}-k\right)
\frac{2k}{\pi}dk,$$
which is continuous with respect to $(x,y)$ in the usual sense (then
in the space $L^2_y([0,X])$ too, with polynomial estimation with
respect to $w$).
\bigskip

On the other hand
$$\int_1^{\infty}\frac{\sin{kx}}{k}\frac{\sin{ky}}{k}
\frac{w}{k+\sqrt{k^2+w}}\frac{2k}{\pi}dk
=\frac{w}{\pi}\int_1^{\infty}\frac{\cos{k(x-y)}-\cos{k(x+y)}}
{k\left(k+\sqrt{k^2+w}\right)}dk.$$
Now let consider the integral with $\cos{k(x-y)}$ : by integrating by parts
we get
$$\int_1^{\infty}\frac{\cos{k(x-y)}}
{k\left(k+\sqrt{k^2+w}\right)}dk=-\frac{\sin{(x-y)}}{x-y}
\frac{1}{1+\sqrt{1+w}}+\int_1^{\infty}\frac{\sin{k(x-y)}}{x-y}
\frac{2+\sqrt{1+\frac{w}{k^2}}+\frac{1}{\sqrt{1+\frac{w}{k^2}}}}
{k^3\left(1+\sqrt{1+\frac{w}{k^2}}\right)^2}dk.$$
The first term is clearly continuously differentiable with respect to
$(x,y)$ with derivatives of polynomial kind with respect to $w$. Now
assume that $x>y$ and differentiate under the integral
to get
\begin{eqnarray*}
\lefteqn{\int_1^{\infty}\frac{k(x-y)\cos{k(x-y)}-\sin{k(x-y)}}{(x-y)^2}
\frac{2+\sqrt{1+\frac{w}{k^2}}+\frac{1}{\sqrt{1+\frac{w}{k^2}}}}
{k^3\left(1+\sqrt{1+\frac{w}{k^2}}\right)^2}dk}\\
& & 
=\left(\int_{x-y}^1+\int_1^{\infty}\right)\frac{k\cos{k}-\sin{k}}{k^3}
\,\frac{2+\sqrt{1+\frac{w}{k^2}(x-y)^2}
+\frac{1}{\sqrt{1+\frac{w}{k^2}(x-y)^2}}}
{\left(1+\sqrt{1+\frac{w}{k^2}(x-y)^2}\right)^2}dk.
\end{eqnarray*}
The first integral is clearly continuous with respect to $(x,y)$ since
it is an absolutely convergent integral of a continuous
function, and can be extended to $y\leq x$. The second is even an
absolutely convergent integral (since $x\leq X$) of a continuous
function which can be extended to $y\leq x$. Moreover
the derivative is still of polynomial kind with respect to $w$.

Now if $x<y$ the derivative of the integral is
$$\left(\int_{y-x}^1+\int_1^{\infty}\right)\frac{k\cos{k}-\sin{k}}{k^3}
\;\frac{2+\sqrt{1+\frac{w}{k^2}(y-x)^2}
+\frac{1}{\sqrt{1+\frac{w}{k^2}(y-x)^2}}}
{\left(1+\sqrt{1+\frac{w}{k^2}(y-x)^2}\right)^2}\,dk,$$
which is continuous with respect to $(x,y)$ with  continuous extension
on $\{x\leq y\}$ (and of polynomial kind with respect to $w$).

In the same way the integral with $\cos{k(x+y)}$ is
$$\int_1^{\infty}\frac{\cos{k(x+y)}}
{k\left(k+\sqrt{k^2+w}\right)}dk=-\frac{\sin{(x+y)}}{x+y}
\frac{1}{1+\sqrt{1+w}}+\int_1^{\infty}\frac{\sin{k(x+y)}}{x+y}
\frac{2+\sqrt{1+\frac{w}{k^2}}+\frac{1}{\sqrt{1+\frac{w}{k^2}}}}
{k^3\left(1+\sqrt{1+\frac{w}{k^2}}\right)^2}dk,$$
and the derivative of the second integral is
$$\left(\int_{x+y}^1+\int_1^{\infty}\right)\frac{k\cos{k}-\sin{k}}{k^3}
\,\frac{2+\sqrt{1+\frac{w}{k^2}(x+y)^2}
+\frac{1}{\sqrt{1+\frac{w}{k^2}(x+y)^2}}}
{\left(1+\sqrt{1+\frac{w}{k^2}(x+y)^2}\right)^2}dk,$$
which is continuous on $\mathbb{R}^+\times\mathbb{R}^+$ and of
polynomial kind with respect to $w$.

Notice that $\frac{\partial\Phi}{\partial x}$ does not exist on
$\{x=y\}$ since the limits from each side are different : indeed
\begin{eqnarray*}
\lim_{(x-y)\rightarrow0^+}\frac{\partial\Phi}{\partial x}(x,y,w)
-\lim_{(x-y)\rightarrow0^-}\frac{\partial\Phi}{\partial x}(x,y,w)
=2\int_0^{\infty}\frac{k\cos{k}-\sin{k}}{k^3}dk\neq0.
\end{eqnarray*}
Nevertheless the application 
$$x\mapsto\Phi(x,y,w)\in L^2_y([0,X])$$
is continuously differentiable :
\begin{eqnarray*}
\lefteqn{\left\|\frac{\Phi(x+h,y,w)-\Phi(x,y,w)}{h}
-\frac{\partial\Phi}{\partial x}(x,y,w)\right\|^2_{L^2_y([0,X])}=}\\
& & =\int_{y\neq x}\left|\frac{\Phi(x+h,y,w)-\Phi(x,y,w)}{h}
-\frac{\partial\Phi}{\partial x}(x,y,w)\right|^2dy,
\end{eqnarray*}
and for all $y\neq x$
$$\frac{\Phi(x+h,y,w)-\Phi(x,y,w)}{h}
\xrightarrow[h\rightarrow0]{}\frac{\partial\Phi}{\partial x}(x,y,w),$$
with domination in $L^2_y([0,X])$ (since $\frac{\partial\Phi}{\partial x}$ 
can be continuously extended on $\{x\geq y\}$ and $\{x\leq y\}$
although the limits do not coincide) then
$$\left\|\frac{\Phi(x+h,y,w)-\Phi(x,y,w)}{h}
-\frac{\partial\Phi}{\partial x}(x,y,w)\right\|_{L^2_y([0,X])}
\xrightarrow[h\rightarrow0]{}0.$$
By the same argument one proves that the derivative is continuous :
$$\int_{y\neq x}\left|\frac{\partial\Phi}{\partial x}(x+h,y,w)
-\frac{\partial\Phi}{\partial x}(x,y,w)\right|^2dy
\xrightarrow[h\rightarrow0]{}0.$$

At last
$\sup_{x\leq X}\left\|\frac{\partial\Phi}{\partial x}(x,y,w)
\right\|_{L^2_y([0,X])}$
is of polynomial kind with respect to $w$.
\bigskip

Now we have to prove the last assertion : for all $x>0$
\begin{eqnarray*}
\Phi(x,x,w) & = & \frac{2w}{\pi}\int_0^{\infty}\frac{\sin^2{kx}}
{k\left(k+\sqrt{k^2+w}\right)}dk\\
& = & \frac{2wx}{\pi}\int_0^{\infty}
\frac{\sin^2{k}}{k\left(k+\sqrt{k^2+wx^2}\right)}dk\\
\end{eqnarray*}
(the equality is still true for $x=0$). One can differentiate under
the integral to get
\begin{eqnarray*}
\frac{d}{dx}\Phi(x,x,w) & = & \frac{2w}{\pi}\int_0^{\infty}
\frac{\sin^2{k}}{k\left(k+\sqrt{k^2+wx^2}\right)}dk\\
& & -\frac{2w^2x^2}{\pi}\int_0^{\infty}
\frac{\sin^2{k}}{k\sqrt{k^2+wx^2}\left(k+\sqrt{k^2+wx^2}\right)^2}dk,\\
\end{eqnarray*}
which is continuous with respect to $x\geq0$ and the estimate follows.

\end{proof}

Next we prove the existence and uniqueness of the solution $A$ :

\begin{lemma}\label{solexiste}

For all $x>0$ and $w\in W$ the equation~\ref{equint} has an unique
solution $A(x,y,w)$ for almost all $y\leq x$. Moreover $A(x,y,w)$ is
holomorphic with respect to $w$ as a vector-valued function in
$L^2_y([0,x])$.

\end{lemma}

\begin{proof}

First $x$ and $w$ being fixed, consider the complex Hilbert space
$L^2([0,x])$ with the associate inner product
$$<f,g>=\int_0^x\overline{f(s)}g(s)ds\,.$$
The resolution of equation~\ref{equint} is equivalent to research the
solutions $h(y)\in L^2([0,x])$ such that
$$\left[(Id+K_{x,w})(h)\right](y)=-\Phi(x,y,w),$$
where $K_{x,w}$ is the integral operator of $L^2([0,x])$ :
$h\mapsto\int_0^xh(s)\Phi(s,y,w)ds$ ($K_{x,w}$ is effectively an
operator since by lemma~\ref{phi} $\Phi$ is continuous).

Next the operator $K_{x,w}$ being defined by $\Phi$ (which is continuous
with respect to $y$ and holomorphic with respect to $w$) is compact
and holomorphic (in the Banach space of operators
$\mathcal{E}_{L^2([0,x])}$ with the associate norm) on the domain
$W$. By the analytic 
Fredholm theorem (see~\cite{reed-simon}), either $(Id+K_{x,w})^{-1}$
exists for no $w\in W$, or $(Id+K_{x,w})^{-1}$ exists and is
holomorphic on $W\setminus S$, where $S$ is a discret subset of $W$ ;
in this case for all $w\in S$ the equation $(Id+K_{x,w})(h)=0$ has a nonzero
solution in $L^2([0,x])$.

The first case is not possible since for $w$ small enough
$\Phi(x,\cdot,w)$ is small enough for all $x,y\in[0,X]$ (see the
beginning of the proof of lemma~\ref{phi}), then
$(Id+K_{x,w})^{-1}$ can be constructed by successive
approximations. In fact the inverse operator exists for all $w\in W$
since $\ker(Id+K_{x,w})=\{0\}$. Indeed let be 
$h\in L^2([0,x])$ such that (for almost all $y\in[0,x]$)
$$h(y)+\int_0^xh(s)\Phi(s,y,w)ds=0.$$
Since
\begin{eqnarray*}
\Re e<h,\int_0^xh(s)\Phi(s,\cdot,w)ds> & = & 
\Re e\int_0^x\overline{h(y)}dy\int_0^xh(s)ds
\int_0^{\infty}\frac{\sin{ks}}{k}\frac{\sin{ky}}{k}
\left(\sqrt{k^2+w}-k\right)\frac{2k}{\pi}dk\\
& = & \int_0^{\infty}\Re e\left(\sqrt{k^2+w}-k\right)\frac{2k}{\pi}dk
\left|\int_0^xh(y)\frac{\sin{ky}}{k}dy\right|^2\\
& \geq & \int_0^{\infty}\left(\sqrt{k^2+\Re e\,w}-k\right)\frac{2k}{\pi}dk
\left|\int_0^xh(y)\frac{\sin{ky}}{k}dy\right|^2\\
& \geq & 0
\end{eqnarray*}
(the permutation of integrals is possible because 
$h\in L^2([0,x])\subset L^1([0,x])$ ), it follows that
\begin{eqnarray*}
0 & = & \Re e\left(<h,h+\int_0^xh(s)\Phi(s,\cdot,w)ds>\right)\\
& \geq & \int_0^x|h(y)|^2dy
\end{eqnarray*}
thus $h=0$. The operator $(Id+K_{x,w})^{-1}$ exists for all $w\in W$
and is holomorphic on $W$ as an operator-valued of $L^2([0,x])$.

It follows that the equation~\ref{equint} has an unique solution
$A(x,\cdot,w)\in L^2([0,x])$ given by
$$A(x,\cdot,w)=(Id+K_{x,w})^{-1}(-\Phi(x,\cdot,w)),$$
equality being considered in the space $L^2([0,x])$ then for almost all
$y\in[0,x]$. At last $A(x,\cdot,w)$ is holomorphic as a vector-valued
function in $L^2([0,x])$ since $(Id+K_{x,w})^{-1}$ is a holomorphic
operator, and $-\Phi(x,y,w)$ is holomorphic on the usual
sense then even on the space $L^2([0,x])$.

\end{proof}

Before proving the regularity of $A$ we need the following lemmas :

\begin{lemma}\label{operator1}

Fix $x_0\in[0,X]$ and a neighbourhood $V_0$ of $x_0$, and consider
the operator 
$L_{x_0,w}=(Id+K_{x_0,w})^{-1}$. Then for all continuous function $f(x,y)$
on $V_0\times[0,X]$ such that the 
application 
$$x\in V_0\mapsto f(x,y)\in L^2_y([0,X])$$ 
is continuously
differentiable, the image $L_{x_0,w}(f(x,\cdot))$ satisfies the same
properties as $f$ and 
$$\frac{\partial}{\partial x}L_{x_0,w}(f(x,\cdot))
=L_{x_0,w}\left(\frac{\partial f}{\partial x}(x,\cdot)\right).$$
Moreover there are the following estimates :
$$\sup_{x\in V_0,y\leq X}|L_{x_0,w}(f(x,\cdot)(y)|\leq
\left(1+x_0\|\Phi(\cdot,w)\|_{\infty}\|L_{x_0,w}\|_{L^2([0,x_0])}\right)
\|f\|_{\infty}$$
and
$$\sup_{x\in V_0}\left\|\frac{\partial}{\partial x}L_{x_0,w}
(f(x,\cdot))(y)\right\|_{L^2_y([0,X])}
\leq\left(1+X\|\Phi(\cdot,w)\|_{\infty}\|L_{x_0,w}\|_{L^2([0,x_0])}\right)
\sup_{x\in V_0}
\left\|\frac{\partial f}{\partial x}(x,\cdot)\right\|_{L^2([0,X])}.$$

\end{lemma}

\begin{proof}

First the element $L_{x_0,w}(f(x,\cdot))$ is well-defined, and
since $f(x,\cdot)=(Id+K_{x_0,w})(L_{x_0,w}(f(x,\cdot))$ (equality in
$L^2([0,x_0])$ ) then for almost all $y\in[0,x_0]$
$$L_{x_0,w}(f(x,\cdot))(y)=f(x,y)
-\int_0^{x_0}L_{x_0,w}(f(x,\cdot))(s)\Phi(s,y,w)ds.$$
The integral does not depend on the choice of the representant of
$L_{x_0,w}(f(x,\cdot))$ and gives a continuous function with respect
to $y$. Since the right member is still defined for $y\in[0,X]$ it
follows that the representant of $L_{x_0,w}(f(x,\cdot))$ can be chosen
as a continuous function wich can be extended on $[0,X]$ (and the
equality will be true for all 
$y\in[0,X]$ ).

Since $f$ is continuous with respect to $(x,y)$ the application
$x\mapsto f(x,\cdot)\in L^2_y([0,x_0])$ is continuous, so is
$L_{x_0,w}(f(x,\cdot))$. It follows that the above
integral is continuous 
with respect to $(x,y)\in V_0\times[0,X]$, as the function
$L_{x_0,w}(f(x,\cdot))(y)$, and using the Cauchy-Schwarz inequality :
$$|L_{x_0,w}(f(x,\cdot)(y)|\leq|f(x,y)|+\|L_{x_0,w}\|_{L^2([0,x_0])}
\|f(x,s)\|_{L^2_s([0,x_0])}\|\Phi(s,y,w)\|_{L^2_s([0,x_0])},$$
then
$$\|L_{x_0,w}(f(x,\cdot)(y)\|_{\infty}\leq\|f\|_{\infty}
+\|L_{x_0,w}\|_{L^2([0,x_0])}\sqrt{x_0}\|f\|_{\infty}\,
\sqrt{x_0}\|\Phi(\cdot,w)\|_{\infty},$$
which proves the first estimate.
\bigskip

Next, the application $x\mapsto f(x,\cdot)\in L^2_y([0,X])$ being
continuously differentiable and $L_{x_0,w}$ an operator of
$L^2([0,x_0])$, it follows that the application
$$x\mapsto L_{x_0,w}(f(x,\cdot))\in L^2_y([0,x_0])$$
is continuously differentiable too and
$$\frac{\partial}{\partial x}L_{x_0,w}(f(x,\cdot))
=L_{x_0,w}\left(\frac{\partial f}{\partial x}(x,\cdot)\right).$$ 
In order to extend the property on
$[0,X]$ one can differentiate (in $L^2_y([0,x_0])$ ) the above formula
to get
$$L_{x_0,w}\left(\frac{\partial f}{\partial x}(x,\cdot)\right)(y)
=\frac{\partial f}{\partial x}(x,y)
-\int_0^{x_0}L_{x_0,w}\left(\frac{\partial f}{\partial x}(x,\cdot)\right)(s)
\,\Phi(s,y,w)ds,$$
which is still well-defined if $y\in[0,X]$ and gives a function in
$L^2_y([0,X])$ continuous with respect to $x\in V_0$. 

At last the second estimate follows since for all $x\in V_0$,
\begin{eqnarray*}
\left\|\frac{\partial}{\partial x}L_{x_0,w}
(f(x,\cdot))(y)\right\|_{L^2_y([0,X])} & \leq & 
\left\|\frac{\partial f}{\partial x}(x,y)\right\|_{L^2_y([0,X])}\\
& & +\left\|L_{x_0,w}\left(\frac{\partial f}{\partial x}(x,\cdot)\right)(s)
\right\|_{L^2_s([0,x_0])}\|\Phi(s,y,w)\|_{L^2([0,x_0]\times[0,X])}\\
& \leq & \left\|\frac{\partial f}{\partial x}(x,y)\right\|_{L^2_y([0,X])}\\
& & +\|L_{x_0,w}\|_{L^2([0,x_0])}
\left\|\frac{\partial f}{\partial x}(x,s)\right\|_{L^2_s([0,X])}
\sqrt{x_0X}\|\Phi(\cdot,w)\|_{\infty}.
\end{eqnarray*}

\end{proof}

\begin{lemma}\label{operator2}

The assertion is still true if we consider the operator
$H_{x,x_0,w}$ defined by 
$$H_{x,x_0,w}:f(x,\cdot)\mapsto
\left(y\in[0,X]\mapsto\int_{x_0}^xf(x,s)\Phi(s,y,w)ds\right)$$
with derivative
$$x\in V_0\mapsto f(x,x)\Phi(x,y,w)
+\int_{x_0}^x\frac{\partial f}{\partial x}(x,s)\Phi(s,y,w)ds,$$
and the following estimates (take $V_0=[x_0-\eta,x_0+\eta]$ ) :
$$\sup_{x\in V_0,y\leq X}|H_{x,x_0,w}(f(x,\cdot))(y)|\leq
\eta\,\|\Phi(\cdot,w)\|_{\infty}\|f\|_{\infty},$$
and
\begin{eqnarray*}
\sup_{x\in V_0}\left\|\frac{\partial}{\partial x}
H_{x,x_0,w}(f(x,\cdot))(y)\right\|_{L^2_y([0,X])} & \leq & 
\sqrt{X}\|\Phi(\cdot,w)\|_{\infty}\|f\|_{\infty}\\
& & +\,\sqrt{\eta}\,\sqrt{X}\|\Phi(\cdot,w)\|_{\infty}\sup_{x\in V_0}
\left\|\frac{\partial f}{\partial x}(x,s)\right\|_{L^2_s([0,X])}.
\end{eqnarray*}

\end{lemma}

\begin{proof}

First the function $H_{x,x_0,w}(f(x,\cdot))(y)$ is clearly continuous
on $V_0\times[0,X]$ and the first estimation follows.

Next we will prove a stronger result : the application 
$x\mapsto H_{x,x_0,w}(f(x,\cdot)\in C^0_y([0,X])$ is continuously
differentiable (i.e. with respect to the uniform topology). Indeed
for all $x\in V_0$ and $y\in[0,X]$
$$\left|\frac{\Delta}{\Delta x}\left(\int_{x_0}^xf(x,s)\Phi(s,y,w)ds\right)
-f(x,x)\Phi(x,y,w)-\int_{x_0}^x\frac{\partial f}
{\partial x}(x,s)\Phi(s,y,w)ds\right|\leq$$
\begin{eqnarray*}
& \leq & \left|\frac{1}{h}\int_{x}^{x+h}f(x+h,s)\Phi(s,y,w)ds
-f(x,x)\Phi(x,y,w)\right|\\
& & +\left|\int_{x_0}^x\left(\frac{f(x+h,s)
-f(x,s)}{h}-\frac{\partial f}{\partial x}(x,s)\right)\Phi(s,y,w)ds\right|\\
& \leq & \|f(x+h,s)\Phi(s,y,w)-f(x,x)\Phi(x,y,w)\|_{\infty,s\in[x,x+h]}\\
& & +\left\|\frac{f(x+h,s)-f(x,s)}{h}-\frac{\partial f}{\partial x}(x,s)
\right\|_{L^2_s([0,X])}\|\Phi(s,y,w)\|_{L^2_s([0,X])}\,,
\end{eqnarray*}
which tends to $0$ uniformly on $y\in[0,X]$ (by uniform continuity of
$f$ and $\Phi$, see lemma~\ref{phi}).
Since the application
$x\mapsto\frac{\partial f}{\partial x}(x,s)\in L^2_s([0,X])$ is
continuous, it follows that the derivative
$$x\in V_0\mapsto f(x,x)\Phi(x,y)
+\int_{x_0}^x\frac{\partial f}{\partial x}(x,s)\Phi(s,y)ds\in C^0_y([0,X])$$
is continuous, and the second estimate can be deduced : for all
$x\in V_0$,
\begin{eqnarray*}
\left\|\frac{\partial}{\partial x}
H_{x,x_0,w}(f(x,\cdot))(y)\right\|_{L^2_y([0,X])} & \leq &
\sqrt{X}|f(x,x)|\,\|\Phi(x,\cdot,w)\|_{\infty}\\
& & +\,
\sqrt{X}\left\|\frac{\partial f}{\partial x}(x,s)\right\|_{L^2_s([x_0,x])}
\sqrt{|x-x_0|}\|\Phi(\cdot,w)\|_{\infty}.
\end{eqnarray*}

\end{proof}

Now the regularity of $A$ can be proved :

\begin{lemma}\label{regular}

The function $A(x,y,w)$ is continuous with respect to $(x,y)\in\Delta$,
 holomorphic with respect to $w\in W$ (in the usual sense) and such
that the application $x\mapsto A(x,y,w)\in L^2_y([0,x])$
is continuously differentiable. Particularly the equation~\ref{equint} is
satisfied for all $(x,y)\in\Delta$ and $w\in W$ (and not only for
almost $y\leq x$).

\end{lemma}

\begin{proof}

The regularity being a local property, let fix $x_0\in\mathbb{R}^+$ and a
neighbourhood $V_0=[x_0-\eta,x_0+\eta]$. For all $x\in V_0$ and almost
$y$ with
$0\leq y\leq x$, the equation~\ref{equint} is equivalent to 
$$(Id+K_{x,w})(A(x,\cdot,w))(y)=-\Phi(x,y,w).$$
Writing $K_{x,w}=K_{x_0,w}+(K_{x,w}-K_{x_0,w})
=K_{x_0,w}+H_{x,x_0,w}$ and applying
$L_{x_0,w}$ (by lemma~\ref{solexiste}), the equation becomes
$$A(x,y,w)+[L_{x_0,w}\circ H_{x,x_0,w}](A(x,\cdot,w))(y)
=L_{x_0,w}(-\Phi(x,\cdot,w))(y).$$
Now we can solve this integral equation by successive approximations
by setting for $x\in V_0$ :
$$
\begin{cases}
A_0(x,y,w)=-L_{x_0,w}(\Phi(x,\cdot,w))(y),\\
A_{n+1}(x,y,w)=-[L_{x_0,w}\circ H_{x,x_0,w}](A_n(x,\cdot,w))(y),
\text{ for all }n\geq0.
\end{cases}
$$
Assume that the function $\tilde{A}(x,y,w):=\sum_{n\geq0}A_n(x,y,w)$ is
well-defined and continuous for all $x-x_0$ small enough with $x\leq
X$ and $y\leq X$ ($X$ being fixed) and such that
$x\mapsto\tilde{A}(x,y,w)\in L^2_y([0,X])$ is
continuously differentiable. Then by uniqueness of the solution of
equation~\ref{equint} it will follow that
$x\mapsto A(x,y,w)\in L^2_y([0,x])$ is continuously differentiable :
for all $x_0$, $A(x,y,w)$ can be extended on
$[x_0-\eta,x_0+\eta]\times[0,X]$ by a continuously differentiable
function (which can depend on $x_0$). Moreover the
equation~\ref{equint} will be true for all $(x,y,w)$ and $A$
holomorphic in the usual sense (indeed, by
using the Cauchy formula in $L^2_y([0,X])$, there is equality for
almost all $y$ of continuous functions with respect to $y$).
\bigskip

First we claim that for all $n\geq0$, $A_n(x,y,w)$ is continuous on
$V_0\times[0,X]$ and such that the application 
$x\in V_0\mapsto A_n(x,y,w)\in L^2_y([0,X])$ is continuously differentiable.
This is true for $n=0$ : indeed, by lemma~\ref{phi} these properties
are satisfied by $\Phi$, and by
lemma~\ref{operator1} this is still true for $A_0(x,y,w)$.

Now if the property is true with $A_n$, by lemmas~\ref{operator1}
and~\ref{operator2} it is still true with
$A_{n+1}(x,y,w)=-[L_{x_0,w}\circ H_{x,x_0,w}](A_n(x,\cdot,w))(y)$. 
\bigskip

Next, in order to prove that $\widetilde{A}$ is continuous on
$V_0\times[0,X]$ and
$x\in V_0\mapsto A(x,y,w)\in L^2_y([0,X])$ is continuously
differentiable, it is sufficient to prove that the following power series
$$\sum_{n\geq0}\|A_n(\cdot,w)\|_{\infty}\text{ and }
\sum_{n\geq0}\,\sup_{x\in V_0}\left\|\frac{\partial A_n}{\partial x}
(x,y,w)\right\|_{L^2_y([0,X])}$$
are convergent. For all $n\geq0$, by lemmas~\ref{operator1}
and~\ref{operator2}
\begin{eqnarray*}
\|A_{n+1}(\cdot,w)\|_{\infty} & \leq & 
\left(1+x_0\|\Phi(\cdot,w)\|_{\infty}\|L_{x_0,w}\|_{L^2([0,x_0])}\right)
\sup_{x\in V_0,y\leq X}|H_{x,x_0,w}(A_n(x,\cdot,w))(y)|\\
& \leq &
\eta\left(1+x_0\|\Phi(\cdot,w)\|_{\infty}\|L_{x_0,w}\|_{L^2([0,x_0])}\right)
\|\Phi(\cdot,w)\|_{\infty}\|A_n(\cdot,w)\|_{\infty}\\
& \leq & \frac{\|A_n(\cdot,w)\|_{\infty}}{4}\,,
\end{eqnarray*}
for $\eta$ small enough (and depending on $x_0,X,w$). It follows that
for all $n\geq0$
$$\|A_n(\cdot,w)\|_{\infty}\leq\frac{\|A_0(\cdot,w)\|_{\infty}}{4^n}$$
and the convergence of the first power serie is proved.

Now by lemmas~\ref{operator1} and~\ref{operator2}, for all $n\geq2$
and uniformly on $x\in V_0$,
\begin{eqnarray*}
\left\|\frac{\partial A_{n+1}}{\partial x}
(x,y,w)\right\|_{L^2_y([0,X])}
& \leq & \left(1+X\|\Phi(\cdot,w)\|_{\infty}
\|L_{x_0,w}\|_{L^2([0,x_0])}\right)\left\|\frac{\partial}{\partial x}
H_{x,x_0,w}(x,y,w)\right\|_{L^2_y([0,X])}\\
& \leq & \left(1+X\|\Phi(\cdot,w)\|_{\infty}
\|L_{x_0,w}\|_{L^2([0,x_0])}\right)\sqrt{X}\|\Phi(\cdot,w)\|_{\infty}\\
& & \times\left(\|A_n(\cdot,w)\|_{\infty}+\sqrt{\eta}\sup_{x\in V_0}
\left\|\frac{\partial A_n}{\partial x}
(x,y,w)\right\|_{L^2_y([0,X])}\right).\\
\end{eqnarray*}
Set
$$C_0(X,x_0,w)
=\max\left\{\|A_0(\cdot,w)\|_{\infty},\;\sqrt{X}\|\Phi(\cdot,w)\|_{\infty}
\left(1+X\|\Phi(\cdot,w)\|_{\infty}
\|L_{x_0,w}\|_{L^2([0,x_0])}\right)\right\},$$
assume that $\sqrt{\eta}\leq\frac{1}{4}$ 
and choose an integer $N_0=N_0(X,x_0,w)$ such that for all $n\geq N_0$
$$\frac{n+2}{\frac{n+1}{2}+\frac{1}{2^{n-1}}}\geq C_0\,.$$
Let consider a constant $C_1=C_1(X,x_0,w)$ large enough such that
$$C_1\geq\max\left\{C_0,\;2^{N_0}\sup_{x\in V_0}
\left\|\frac{\partial A_{N_0}}{\partial x}
(x,y,w)\right\|_{L^2_y([0,X])}\right\}.$$
Then ones proves that for all $n\geq N_0$
$$\sup_{x\in V_0}\left\|\frac{\partial A_n}{\partial x}
(x,y,w)\right\|_{L^2_y([0,X])}\leq C_1\frac{n+1}{2^n}\,.$$
By construction it is true for $n=N_0$. Assume that it is true for $n$
then from above (and since 
$\|A_n(\cdot,w)\|_{\infty}\leq\frac{\|A_0(\cdot,w)\|_{\infty}}{4^n}
\leq\frac{C_0}{4^n}$ ) it follows that
\begin{eqnarray*}
\sup_{x\in V_0}\left\|\frac{\partial A_{n+1}}{\partial x}
(x,y,w)\right\|_{L^2_y([0,X])} & \leq &
C_0\left(\frac{C_0}{4^n}+\frac{C_1}{4}\frac{n+1}{2^n}\right)\\
& \leq & C_0C_1\left(\frac{1}{4^n}+\frac{n+1}{2^{n+2}}\right)\\
& \leq & C_1\frac{n+2}{2^{n+1}},
\end{eqnarray*}
by construction of $N_0$.

The recurrence is proved and it follows that the power serie
$$\sum_{n\geq0}\sup_{x\in V_0}\left\|\frac{\partial A_n}{\partial x}
(x,y,w)\right\|_{L^2_y([0,X])}$$
is convergent and the proof is complete.

\end{proof}

Before proving the estimates of $A$ we need the following lemma which
is the continuity of the inverse integral operator : 

\begin{lemma}\label{lemma3}

For all $x>0$ and $w\in W$, for all $h\in L^2([0,x])$,
$$\|h\|_{L^2([0,x])}\leq
\left\|h(y)+\int_0^xh(s)\Phi(s,y,w)ds\right\|_{L^2_y([0,x])}.$$

\end{lemma}

\begin{proof}

First
$$\left\|h(y)+\int_0^xh(s)\Phi(s,y,w)ds\right\|_{L^2_y([0,x])}^2\geq
\int_0^x|h(y)|^2dy
+2\Re e\int_0^x\overline{h(y)}dy\int_0^xh(s)\Phi(s,y,w)ds,$$
and it has been proved (see proof of lemma~\ref{solexiste}) that for
all $w\in W$,
$$\Re e\int_0^x\overline{h(y)}dy\int_0^xh(s)\Phi(s,y,w)ds\geq0,$$
the inequality follows.

\end{proof}

The proof of proposition~\ref{solution} can be complete thanks to the
last lemma :

\begin{lemma}\label{estimations}

The function $A(x,y,w)$ is of poynomial kind with respect to $w$ : for all
$X\geq1$ and $w\in W$,
$$\sup_{y\leq x\leq X}|A(x,y,w)|,\;
\sup_{y\leq x\leq X}\left\|\frac{\partial A}{\partial x}
(x,y,w)\right\|_{L^2_y([0,X])},\;
\sup_{x\leq X}\left|\frac{d}{dx}A(x,x,w)\right|
\leq C(X)(1+|w|)^{\alpha}\,.$$
($\alpha$ does not depend on $X$)

\end{lemma}

\begin{proof}

During the proof we will use the same notation for different constants
$C(X)$ and $\alpha$.
\bigskip

First by definition of $A$ and lemma~\ref{lemma3}
\begin{eqnarray*}
\|A(x,y,w)\|_{L^2_y([0,x])} & \leq & \left\|A(x,y,w)
+\int_0^xA(x,s,w)\Phi(s,y,w)ds\right\|_{L^2_y([0,x])}\\
& = & \|\Phi(x,y,w)\|_{L^2_y([0,x])}\\
& \leq & \sqrt{x}\|\Phi(\cdot,w)\|_{\infty}\\
& \leq & C(X)(1+|w|)^{\alpha},
\end{eqnarray*}
the last inequality coming from lemma~\ref{phi}. It follows that for
all $0\leq y\leq x\leq X$
\begin{eqnarray*}
|A(x,y,w)| & = & \left|\int_0^xA(x,s,w)\Phi(s,y,w)ds+\Phi(x,y,w)\right|\\
& \leq & \|A(x,s,w)\|_{L^2_s([0,x])}\|\Phi(s,y,w)\|_{L^2_s([0,x])}
+\|\Phi(\cdot,w)\|_{\infty}
\end{eqnarray*}
thus
$$\|A(\cdot,w)\|_{\infty}\leq C(X)(1+|w|)^{\alpha},$$
this proves the first estimate.
\bigskip

Next by differentiating the equation~\ref{equint} with respect to $x$
(which is possible by lemma~\ref{regular}) we get for all $x\leq X$
$$\frac{\partial A}{\partial x}(x,y,w)
+\int_0^x\frac{\partial A}{\partial x}(x,s,w)
\Phi(s,y,w)ds+A(x,x,w)\Phi(x,y,w)
+\frac{\partial\Phi}{\partial x}(x,y,w)=0,$$
thus by lemma~\ref{lemma3}
$$\left\|\frac{\partial A}{\partial x}(x,y,w)\right\|_{L^2_y([0,x])}
\leq\left\|A(x,x,w)\Phi(x,y,w)+\frac{\partial\Phi}{\partial x}
(x,y,w)\right\|_{L^2_y([0,x])},$$
and by lemma~\ref{phi}
$$\sup_{x\leq X}
\left\|\frac{\partial A}{\partial x}(x,y,w)\right\|_{L^2_y([0,x])}
\leq C(X)(1+|w|)^{\alpha}.$$
\bigskip

In order to prove the last estimate, take $y=x$ in the
equation~\ref{equint} and differentiate (which is possible since by
lemmas~\ref{phi}, $\frac{\partial\Phi}{\partial x}$ exist in
$L^2_y([0,X])$~) to get
\begin{eqnarray*}
\lefteqn{\frac{d}{dx}A(x,x,w)
+\int_0^x\frac{\partial A}{\partial x}(x,s,w)\Phi(s,x,w)ds}\\
& & +A(x,x,w)\Phi(x,x,w)
+\int_0^xA(x,s,w)\frac{\partial\Phi}{\partial x}(x,s,w)ds
+\frac{d}{dx}\Phi(x,x,w)=0,
\end{eqnarray*}
since $\Phi(x,s,w)=\Phi(s,x,w)$.
It follows that for all $x\leq X$
\begin{eqnarray*}
\left|\frac{d}{dx}A(x,x,w)\right| & \leq &
|A(x,x,w)\Phi(x,x,w)|
+\;\left\|\frac{\partial A}{\partial x}(x,s,w)\right\|_{L^2_s([0,x])}
\left\|\Phi(s,x,w)\right\|_{L^2_s([0,x])}\\
& & +\left\|A(x,s,w)\right\|_{L^2_s([0,x])}
\left\|\frac{\partial\Phi}{\partial x}(s,x,w)\right\|_{L^2_s([0,x])}\;
+\left|\frac{d}{dx}\Phi(x,x,w)\right|,
\end{eqnarray*}
thus by lemma~\ref{phi} and from above
$$\sup_{x\leq X}\left|\frac{d}{dx}A(x,x,w)\right|\leq
C(X)(1+|w|)^{\alpha}.$$

At last the exponent $\alpha$ does not depend on $X$ since it is true
with $\Phi$ (see lemma~\ref{phi}).

\end{proof}

\section{Applications in Sturm-Liouville inverse
  problems}\label{application}

\subsection{Some reminds and motivation of the problem}
We consider here the equation on the half-axis $\mathbb{R}^{+}$
$$-y''(x)-\omega^2Q(x)=\lambda y(x),$$
with the following hypothesis : $Q$ is strictly positive and strictly
decreasing, integrable with $m+1$ locally integrable derivatives which
are polynomially decreasing at infinity. Then we know that there are
$N(\omega)$ (of order $\omega$) eigenvalues $\lambda_j=-\xi_j^2$, with
$N(\omega)$ $L^2$-normed eigenfunctions $\phi_j$ which satisfy
$\phi_j(0)=0$.

Here we deal with an inverse problem : if for any $\omega$ large
enough we know the $N(\omega)$ eigenvalues $\xi_j$ and characteristic
values 
$$C_j=\left(\phi_j'(0)\right)^2,$$
$j=1,\ldots,N(\omega)$, we have to get back the potential
$-\omega^2Q$ on $\mathbb{R}^+$. More generally the aim of inverse
theory is to reconstruct $Q$ from informations of its solutions : we
principally deal with the case where these informations are given by
the Weyl function defined for $\Im m\,k>0$ as
$$j(k)=\frac{\varphi'(0,k)}{\varphi(0,k)},$$
with $\phi$ a $L^2$-integrable solution (and $\lambda=k^2$). We know that
it is a meromorphic function on the half-plane of which the poles are
exactly the eigenvalues $i\xi_j$ and the residues are (modulo
multiplication by $2i\xi_j$) the characteristic values $C_j$. Thus we
can determine the spectral measure $\sigma(d\tau)$ of the potential
$-\omega^2Q$ :
$$\sigma(d\tau)=
\begin{cases}
\sigma_+(d\tau),\;\tau\geq0,\\
\sum_{j=1}^{N(\omega)}C_j\delta(\tau+\xi_j^2),\;\tau<0,
\end{cases}
$$
where $\sigma_+$ is a positive measure with a density function and
$\delta$ the Dirac measure. Thanks to the works of Gelfand and Levitan
we can reconstruct $Q$.

Here we assume that we only know the parameters $\xi_j,\,C_j$ and the
first derivatives of $Q$ at $0$. By the result of G. Henkin
and N. Novikova in~\cite{henkin} (theorem 1, p. 21), one can approach
$Q$ uniformly on any $[0,X]$, with
precision of order 
$$\frac{1}{\omega^m},$$
by a function $Q_{\omega}$ arising from the potential
$q_{\omega}(x)=-\omega^2Q_{\omega}(x)$
associate to the explicit spectral measure $\sigma_{\omega}(d\tau)$
constructed with the $\xi_j$, $C_j$ and $Q^{(s)}(0)$ for
$s=0,\ldots,m$ (see~\cite{henkin}).

For $m=1$ there is an explicit potential
$q^0_{\omega}=-\omega^2Q_{\omega}^0$ defined by (cf~\cite{henkin}, p. 23)
$$Q_{\omega}^0(x)=\frac{2}{\omega^2}\frac{d^2}{dx^2}\ln|\det W_{s,r}(x)|,$$
where
$$W_{s,r}(x)=\frac{2sh(\xi_s+\xi_r)x}{\xi_s+\xi_r}-(1-\delta_{s,r})
\frac{2sh(\xi_s-\xi_r)x}{\xi_s-\xi_r}-\delta_{s,r}\left(2x-\frac{4\xi_r^2}
{C_r}\right),$$
$s,\,r=1,\ldots,N(\omega)$. There is uniformly on any $[0,X]$
$$\left|\int_0^xQ(y)dy-\int_0^xQ_{\omega}^0(x)\right|=
O\left(\frac{\ln\omega}{\sqrt{\omega}}\right)$$
(but the precision could be better than
$\frac{\ln\omega}{\sqrt{\omega}}$, we think that it should be of order
$\frac{1}{\omega^3}$). 
\bigskip

It has been conjectured in~\cite{henkin}, p. 22 that such a formula
with $2N(\omega)$ parameters (the $\xi_j$ and $C_j$) could not
uniformly approach a function in general position with $m$ bounded
derivatives (i.e. the compact set $\Lambda_m$) better than of order
$\frac{1}{\omega^m}$. Otherwise we have a natural question :  can we
find, or at least prove the 
existence of a formula which would get a better approximation ?

By assuming that such a formula can be written as an analytic function
with respect to the parameters $\xi_j$, $C_j$ (and
$Q^{(s)}(0),\;s=0,\ldots,m$), and 
relating with the negatives results which are given in~\ref{rappel},
we are going to give lower bounds for such approximations in order to
get the two following results about the best reconstruction of potentials
(theorems~\ref{optim1} and~\ref{optim2}).

If $p$ is an integer, $\mathcal{Q}_p$ will mean the class of functions $Q$
which are defined on 
$\mathbb{R}^+$, strictly positive, strictly decreasing with $p$
locally integrable derivatives which vanish at $0$, and
with polynomially behavior at infinity as well as its derivatives.
The set $\Lambda_{\mathcal{Q}_p}$ will mean any compact set of
$\mathcal{Q}_p$ such that ${\Lambda_{\mathcal{Q}_p}}_{|[0,1]}$ is of kind
$\Lambda_p([0,1])$ and for any $Q\in\Lambda_{\mathcal{Q}_p}$ the
parameters $\int_0^{\infty}(1+t)\sqrt{Q(t)}dt$ and $a,\,k$ are bounded
(where $\frac{1}{a}\frac{1}{1+x^k}\leq Q(x)\leq\frac{a}{1+x^k}$~). It
follows that the constants which appear to estimate the eigenvalues
and the characteristic values of Sturm-Liouville operator, only depend
on $\Lambda_{\mathcal{Q}_p}$ (see remark~\ref{constcomp}).

\subsection{Result on the case with 2 derivatives}

As application of our negative results on approximation theory
(corollary~\ref{detail}) and positive results above
(see~\cite{henkin}), we will obtain the following theorem : 

\begin{theorem}\label{optim1}

Let consider $\mathcal{Q}_2$ and $\Lambda_{\mathcal{Q}_2}$, and for any
$\omega$ large enough and $Q\in\Lambda_{\mathcal{Q}_2}$ the Sturm-Liouville
operators $-\frac{d^2}{dx^2}-\omega^2Q$.

For any $N$ with $a_1\omega\leq N\leq a_2\omega$,  let
$\psi(x,\zeta)$ be a function defined on
$\mathbb{R}\times\mathbb{C}^N$, of class $C^1$ with respect to $x$ and
which satisfies the conditions of corollary~\ref{inverse2} on any
$[0,X]$ (with respect to $\omega$). At last let
$b(\omega)>0$ be a constant such that $b(\omega)$ and
$\frac{1}{b(\omega)}$ are polynomial at $\omega$.

Then the approximation of
$$\int_0^{\cdot}\Lambda_{\mathcal{Q}_2}:=\left\{\left(x\mapsto\int_0^xQ(t)dt
\right),\;Q\in\Lambda_{\mathcal{Q}_2}\right\},$$
uniformly on any $[0,X]$ with $X\geq1$, by the family
$$\left\{\left(x\mapsto\frac{1}{b(\omega)}\left(\frac{1}{\psi}
\frac{\partial\psi}{\partial x}
\right)(x,\zeta)\right), \,\zeta_j=O(\omega^r),
\,\forall\,j=1,\ldots,N\right\}$$
when $\omega\rightarrow\infty$, cannot be better than of order of
$$\frac{1}{(\omega\ln\omega)^3}\,.$$

In addition we have got an approximation formula such that if
$N(\omega)$ is the number of eigenvalues
$\xi_j$ and characteristic values $C_j$ of the operator
$-\frac{d^2}{dx^2}-\omega^2Q$ and
$$\Psi(x,\zeta)=\det\widetilde{W}_{s,r}(x,\zeta)$$
with
$$\widetilde{W}_{s,r}(x,\zeta)=\frac{2sh(\zeta_r+\zeta_s)x}{\zeta_r+\zeta_s}
-(1-\delta_{s,r})\frac{2sh(\zeta_s-\zeta_r)x}{\zeta_s-\zeta_r}
-\delta_{s,r}\left(2x-\exp\left(\zeta_{r+N(\omega)}\right)\right),$$
$s,\,r=1,\ldots,N(\omega)$,
then the family
$\left\{\frac{2}{\omega^2}\frac{1}{\Psi}\frac{\partial\Psi}{\partial x}
\right\}$
approximates the compact set $\int_0^{\cdot}\Lambda_{\mathcal{Q}_2}$ at
least with the precision of order
$\frac{\ln\omega}{\sqrt{\omega}}$.

Moreover $Q$ being given, such an element
$\zeta(Q)$ can be chosen as
$$\zeta_j(Q)=\xi_j(Q)\text{ and }
\zeta_{j+N(\omega)}(Q)
=\ln\frac{4\xi_j^2(Q)}{C_j(Q)},\,j=1,\ldots,N(\omega).$$

\end{theorem}

There is no analytic formula which can approximate any
given potential (with $2$ derivatives) with a better precision than of order
$\frac{1}{(\omega\ln\omega)^3}$. The explicit approximating formula
given by Gelfand-Levitan-Jost-Kohn gives a positive result with a
precision of order (at least) $\frac{\ln\omega}{\sqrt{\omega}}$ (as we
already indicated we think that it should be of order
$\frac{1}{\omega^3}$ ). We know the negative result for a polynomial
family (see~\cite{vitushkin},~\cite{warren},~\cite{amadeo}), so it is
natural to wonder what will happen if we consider a nonlinear family
with $\omega$ parameters
in order to get a better approximation. This negative result gives
an answer to the question asked in~\cite{henkin}, p. 22.

On the other side it is interesting to see that the explicit formula
in the positive result
has not been specially constructed in the sense of abstract
approximation theory but it comes from mathematical physics.

\begin{proof}

First from definition of $\Lambda_{\mathcal{Q}_2}$ the restriction on
$[0,1]$ of any $Q$ is in (a homothetic of) $\Lambda_2([0,1])$, and
conversely let $h\in\Lambda_2([0,1])$ be strictly positive, strictly
decreasing with $h'(0)=h''(0)=0$, it can be extended on $\mathbb{R}^+$ to a
function $Q_h$ (then a potential $-\omega^2Q_h$) such that
$Q_h\in\Lambda_{\mathcal{Q}_2}$.

One can apply the corollary~\ref{detail} (for $m=2$) with $\psi$ and
$\Lambda_{\mathcal{Q}_2}$ since $N$ and $\omega$ have same order, and
there is a potential $-\omega^2Q_h,\;Q_h>0$, such that
$\int_0^{\cdot}Q_h$ cannot be closer to
$\left\{\frac{1}{b(\omega)}\frac{1}{\psi}\frac{\partial\psi}
{\partial x}(\cdot,\zeta)\right\}$ on
$[0,1]$, than
$\frac{C}{(N\ln N)^3}$ for all $\zeta$ of polynomial size at
$\omega$, thus of order $\frac{1}{(\omega\ln\omega)^3}$ on any $[0,X]$. 
\bigskip

Now notice that for any $Q\in\Lambda_{\mathcal{Q}_2}$ the choice of
$N_Q(\omega)$ is possible to get since there are constants $b_1$ and
$b_2$ (depending only on $\Lambda_{\mathcal{Q}_2}$) such that for all
$Q$
$$b_1\omega\leq N_Q(\omega)\leq b_2\omega.$$
Indeed $Q$ being decreasing, there are the bounds of Calogero
(see~\cite{calogero},~\cite{reed-simon}) :
$$\frac{\omega}{\pi\sqrt{Q(0)}}\int_0^{\infty}Q(x)dx-\frac{1}{2}
\leq N(\omega)\leq\frac{2\omega}{\pi}\int_0^{\infty}\sqrt{Q(x)}dx$$
with $Q$ in a compact subset.
\bigskip

Next the function $\Psi$ fulfills the conditions of
corollary~\ref{inverse2}. Indeed it is a function with $2N(\omega)$
parameters ($N(\omega)$ and $\omega$ being of same order) and entire of
exponential kind :

it is true with $\frac{sh(\zeta_s\pm\zeta_r)x}{\zeta_s\pm\zeta_r}$ on
any $[0,X]$ then as well as $\widetilde{W}_{s,r}$ and each product of
the determinant :
$$\left\|\prod_{j=1}^{N(\omega)}\widetilde{W}_{j,\tau(j)}\right\|_{\infty}
\leq C_1(X)^{N(\omega)}\exp\left(C_2(X)\|\zeta\|_1\right).$$
Since the number of these products is
$N(\omega)!=O\left(\exp\left(N(\omega)^2\right)\right)$, one gets an
estimate of
$\Psi$ as $O\left(\exp\left(\alpha\omega^{\beta}\right)\right)$.

One can deduce the estimates of $\frac{\partial\Psi}{\partial x}$ since
$\Psi$ is as well entire (and of exponential kind) with respect to
the variable $x$. The Cauchy formula applied on the disc
$D(0,X+1)$ and the above estimation give the same upper estimate about
$\frac{\partial\Psi}{\partial x}$ on any $[0,X]$ and for all
$\zeta\in\mathbb{C}^{2N(\omega)}$.

On the other side
$$\det\left(\widetilde{W}_{s,r}\right)(0)^{-1}=
\frac{1}{2}\prod_{j=1}^{N(\omega)}\exp\left(-\zeta_{N(\omega)+j}\right)
\leq\exp(\|\zeta\|_1)=
O\left(e^{\alpha\omega^{\beta}}\right).$$
\bigskip

And the choice of the valuations
$$\zeta_j(Q)=\xi_j\left(-\omega^2Q\right)\text{ and }\zeta_{j+N(\omega)}(Q)=
\ln\frac{4\xi_j^2}{C_j}\left(-\omega^2Q\right),\;j=1,\ldots,N(\omega),$$
is possible to get since $Q\in\Lambda_{\mathcal{Q}_2}$ satisfies the
conditions of proposition~\ref{estimation} : it follows that
$\xi_j\left(-\omega^2Q\right)=O(\omega)$ and
$$\frac{1}{\alpha\exp(\beta\omega^{\gamma})}\leq
\frac{4\xi_j^2}{C_j}\left(-\omega^2Q\right)\leq
\alpha\exp(\beta\omega^{\gamma})$$
then
$\ln\frac{4\xi_j^2}{C_j}\left(-\omega^2Q\right)=
O\left(\omega^{\gamma}\right)$ (and notice that by
remark~\ref{constcomp} the constants depend only on
$\Lambda_{\mathcal{Q}_2}$).

At last this choice of $\zeta(Q)$ gives the effective approximation
of all $\int_0^{\cdot}Q$ on any $[0,X]$ with precision of order of
$\frac{\ln\omega}{\sqrt{\omega}}$ :
indeed this is an application of the theorem~2 given in~\cite{henkin}
p. 22, which claims that uniformly on any $[0,X]$,
\begin{eqnarray*}
\left|\int_0^xQ(y)dy-\frac{1}{\omega^2}\frac{1}{\Psi(x)}\frac{\partial\Psi}
{\partial x}(x,\zeta(Q))\right| & = &
\left|\int_0^xQ(y)dy-\frac{2}{\omega^2}\int_0^x 
\frac{\partial^2}{\partial x^2}\ln\left|\Psi(y,\zeta(Q))\right|dy\right|\\
& = & \left|\int_0^xQ(y)dy-2\int_0^xQ_{\omega}^0(y)dy\right|\\ 
& = & O\left(\frac{\ln\omega}{\sqrt{\omega}}\right)\,.
\end{eqnarray*}
The conditions of regularity about $Q$ are fulfilled in order to apply
the theorem~: $Q$ is strictly positive (as well lower bounded by a
constant only depending on the compact $\Lambda_{\mathcal{Q}_2}$ ) and
in (a homothetic of) $\Lambda_2$ : this condition can replace
the bounded number of intervals of monotonicity on $\mathbb{R}^+$ of
$Q$ and its derivatives.

\end{proof}

\subsection{Result on the case with $m+1$ derivatives}

Now if we consider the  case with $\mathcal{Q}_{m+1}$ of potentials with
$m+1$ locally integrable derivatives which all vanish at $0$, we can
give a spectral measure $\sigma_{\omega}(d\tau)$ close to the one
associate to the operator $-\frac{d^2}{dx^2}-\omega^2Q$. It has the
following form :
$$\sigma_{\omega}(d\tau)=
\begin{cases}
\frac{1}{\pi}\sqrt{\tau+\omega^2Q(0)},\;\tau\geq0,\\
\sum_{j=1}^{N(\omega)}C_j\delta(\tau+\xi_j^2),\;\tau<0\,.
\end{cases}
$$
Let be for $(x,y)\in\Delta=\{0\leq y\leq x\}$
$$\Phi(x,y)=\frac{1}{\pi}\int_0^{\infty}\frac{\sin(x\sqrt{\tau})}
{\sqrt{\tau}}\frac{\sin(y\sqrt{\tau})}{\sqrt{\tau}}\frac{\omega^2Q(0)
\;d\tau}{\sqrt{\tau+\omega^2Q(0)}+\sqrt{\tau}}\,,$$
(absolutely convergent integral for all $x,\,y$) and consider the kernel
$A(x,y)$ solution of the integral equation
$$A(x,y)+\int_0^xA(x,s)\Phi(s,y)ds+\Phi(x,y)\equiv0\,.$$
At last let $q_{\omega}$ be the following potential :
$$q_{\omega}(x)=2\frac{d}{dx}A(x,x)-2\frac{d^2}{dx^2}\ln|\det T(x)|\,,$$
where $T(x)$ is the matrix of order $N(\omega)$ (see~\cite{levitan}) :
\begin{eqnarray*}
\lefteqn{T_{j,k}(x)=\frac{4\xi_j^2}{C_j}\delta_{j,k}\,+}\\
& & +\,4\int_0^x\left(sh(\xi_jt)+\int_0^tA(t,s)sh(\xi_js)ds\right)
\left(sh(\xi_kt)+\int_0^tA(t,s)sh(\xi_ks)ds\right)\,dt\,.
\end{eqnarray*}
Then (~\cite{henkin}, th{\'e}or{\`e}me 1 p. 21) the function
$Q_{\omega}=-\frac{q_{\omega}}{\omega^2}$ approximates all $Q$
uniformly on any $[0,X]$ with precision of order (at least) of
$\frac{1}{\omega^m}$ (and the precision should as well be better, of
order of $\frac{1}{\omega^{m+1}}$ ).
\bigskip

As we see, $Q_{\omega}$ can be written as an analytic function 
$\widetilde{Q}_{\omega}(\zeta,w)$ with
$\zeta\in\mathbb{C}^{2N}$ and $w\in W=\{\Re e\,z>0\}$ ($w$ replaces
the variable $\omega^2Q(0)$ ). More precisely
$$\widetilde{Q}_{\omega}(x,\zeta,w)=\frac{2}{\omega^2}\left(
-\frac{\partial\widetilde{A}}{\partial x}(x,x,w)+
\frac{\partial^2}{\partial x^2}\ln\left|\det\widetilde{T}(x,\zeta,w)
\right|\right)\,,$$
where $\widetilde{A}$ is the solution, defined for all
$0\leq y\leq x$ and $\Re e\,w>0$, of the equation
$$\widetilde{A}(x,y,w)+\int_0^x\widetilde{A}(x,s,w)
\widetilde{\Phi}(s,y,w)ds+\widetilde{\Phi}(x,y,w)=0$$
with
\begin{eqnarray*}
\widetilde{\Phi}(x,y,w) & = & \frac{1}{\pi}\int_0^{\infty}
\frac{\sin(x\sqrt{\tau})}{\sqrt{\tau}}\frac{\sin(y\sqrt{\tau})}{\sqrt{\tau}}
\frac{w\;d\tau}{\sqrt{\tau+w}+\sqrt{\tau}}\\
& = & \int_0^{\infty}
\frac{\sin{kx}}{k}\frac{\sin{ky}}{k}
\frac{w}{k+\sqrt{k^2+w}}\frac{2k}{\pi}dk
\end{eqnarray*}
($\widetilde{\Phi}$ is well defined since $\Re e\,w>0$) and
\begin{eqnarray*}
\widetilde{T}_{j,k}(x,w,\zeta) & = & \exp(\zeta_{N+j})\delta_{j,k}\\
& + & 4\int_0^x\left(sh(\zeta_jt)+\int_0^t\widetilde{A}(t,s,w)
sh(\zeta_js)ds\right)\left(sh(\zeta_kt)+\int_0^t\widetilde{A}(t,s,w)
sh(\zeta_ks)ds\right)dt.
\end{eqnarray*}

Now there is the following theorem :

\begin{theorem}\label{optim2}

Consider in the same way $\mathcal{Q}_{m+1}$,
$\Lambda_{\mathcal{Q}_{m+1}}$ and the Sturm-Liouville
operators $-\frac{d^2}{dx^2}-\omega^2Q$ with
$Q\in\Lambda_{\mathcal{Q}_{m+1}}$ and $\omega$ large enough.

For any $(N,M)$ with $a_1\omega\leq N+M\leq a_2\omega$, let
$\psi(x,\zeta,w)$ (resp. $k(x,\zeta,w)$~) be a function defined on
$\mathbb{R}^+\times\mathbb{C}^N\times W^M$ of class $C^2$
(resp. continuous) with respect to $x$ and which
which satisfies the conditions of corollary~\ref{inverse1} on any
$[0,X]$ with $\Omega_{\omega}=\Omega_{N,M}$. At last
let $b(\omega)>0$ be a constant such that $b(\omega)$ and
$\frac{1}{b(\omega)}$ are polynomial at $\omega$.

Then the approximation of $\Lambda_{\mathcal{Q}_{m+1}}$
uniformly on any $[0,X]$ by the family
$$\left\{\left(x\mapsto\frac{1}{b(\omega)}\left(k(x,\zeta,w)+
\frac{\partial}{\partial x}
\left(\frac{1}{\psi}\frac{\partial\psi}{\partial x}\right)(x,\zeta,w)\right)
\right),\;(\zeta,w)\in\Omega_{\omega}\right\},$$
when $\omega\rightarrow\infty$, cannot be better than of order of
$$\frac{1}{(\omega\ln\omega)^{m+1}}\,.$$

In addition there is an almost optimal approximation formula such that, if
$N(\omega)$ is the number of eigenvalues
$\xi_j$ and characteristic values $C_j$ of the operator
$-\frac{d^2}{dx^2}-\omega^2Q$ (and $M=1$) and
$\left(\widetilde{k},\widetilde{\Psi}\right)$ 
is defined on $\mathbb{R}^+\times\mathbb{C}^{2N(\omega)}\times W$ as
$$\widetilde{k}(x,w)=-\frac{\partial}{\partial x}\widetilde{A}(x,x,w)
\text{ and }\widetilde{\Psi}(x,\zeta,w)=\det\widetilde{T}_{j,k}
(x,\zeta,w)$$
with $\widetilde{A}$ and $\widetilde{T}_{j,k}$ defined as above,
then the family
$$\left\{\frac{2}{\omega^2}\left(\widetilde{k}(x,w)
+\frac{\partial}{\partial x}\left(\frac{1}{\widetilde{\Psi}}
\frac{\partial\widetilde{\Psi}}{\partial x}\right)(x,\zeta,w)\right)
\right\}$$
approximates $\Lambda_{\mathcal{Q}_{m+1}}$ at
least with the precision of order of
$$\frac{1}{\omega^m}.$$

Moreover $Q\in\Lambda_{\mathcal{Q}_{m+1}}$ being given, such an element
$(\zeta(Q),w(Q))$ can be chosen as
$$\zeta_j(Q)=\xi_j(Q),\;\zeta_{j+N(\omega)}(Q)
=\ln\frac{4\xi_j^2(Q)}{C_j(Q)},
\;j=1,\ldots,N(\omega),\text{ and }\;w(Q)=\omega^2Q(0).$$

\end{theorem}

\begin{proof}

The negative result follows on $[0,1]$ by corollary~\ref{detail} since
$N+M$ and $\omega$ have same order. Moreover the restriction on $[0,1]$
of any potential gives a function in $\Lambda_{m+1}$ ; conversely any
function
$h\in\Lambda_{m+1}([0,1])$, strictly decreasing and satisfying
$h'(0)=\cdots=h^{(m+1)}(0)=0$,
can be extended on $\mathbb{R}^+$ by
$Q_h\in\Lambda_{\mathcal{Q}_{m+1}}$. This proves the negative part
with the compact $\Lambda_{\mathcal{Q}_{m+1}}$.

On the other hand the positive result with approximation at order of
$\frac{1}{\omega^m}$ comes from the theorem~1 of~\cite{henkin} (notice
that the condition
$Q\in\Lambda_{\mathcal{Q}_{m+1}}$ can replace the bounded number of
intervals of monotonicity of $Q$ and its derivatives). Then the only
difficulty is to prove that the choice 
of $\left(\widetilde{A},\widetilde{\Psi}\right)$ is possible to get.

First by proposition~\ref{solution} the function 
$$x\mapsto\frac{\partial}{\partial x}\widetilde{A}(x,x,w)$$ 
exists, is continuous with respect to $x$ and holomorphic of
polynomial kind with respect to $w\in W$.

Next each function $\widetilde{T}_{j,k}$ is of class $C^2$ with
respect to $x$ : indeed $\widetilde{A}(x,y,w)$ is continuous with
respect to $(x,y)\in\Delta$ then
$$\frac{\partial\widetilde{T}_{j,k}}{\partial x}(x,\zeta,w)
=4\left(sh(\zeta_jx)+\int_0^x\widetilde{A}(x,s,w)
sh(\zeta_js)ds\right)\left(sh(\zeta_kx)+\int_0^x\widetilde{A}(x,s,w)
sh(\zeta_ks)ds\right).$$
Since the application $x\mapsto\widetilde{A}(x,y,w)\in L^2_y([0,x])$
is continuously differentiable, it follows that for all $j,\,k$
\begin{eqnarray*}
\frac{\partial^2\widetilde{T}_{j,k}}{\partial x^2}(x,\zeta,w) & = &
4\left(\zeta_jch(\zeta_jx)+\widetilde{A}(x,x,w)sh(\zeta_jx)
+\int_0^x\frac{\partial\widetilde{A}}{\partial x}(x,s,w)
sh(\zeta_js)ds\right)\\
& & \times\left(sh(\zeta_kx)+\int_0^x\widetilde{A}(x,s,w)
sh(\zeta_ks)ds\right)\\
& + & 4\left(sh(\zeta_jx)+\int_0^x\widetilde{A}(x,s,w)
sh(\zeta_js)ds\right)\\
& & \times\left(\zeta_jch(\zeta_kx)+\widetilde{A}(x,x,w)sh(\zeta_kx)
+\int_0^x\frac{\partial\widetilde{A}}{\partial x}(x,s,w)
sh(\zeta_ks)ds\right)\\
\end{eqnarray*}
exists and is continuous with respect to $x\in\mathbb{R}^+$. Then
$\widetilde{\Psi}$ is of class $C^2$ with respect to $x$
and the estimates are fulfilled : indeed it is true for each
$\widetilde{T}_{j,k}$ and its two derivatives thanks to
proposition~\ref{solution} hence it is true for the determinant
$\widetilde{\Psi}$ and its derivatives.

On the other hand $\widetilde{k}$ is continuous with respect to $x$
and holomorphic of polynomial kind with respect to all $w\in W$ (and
not only in $\mathbb{R}^+$).

Moreover for all $j,k$ and all
$(\zeta,w)\in\mathbb{C}^{2N(\omega)}\times W$
$$\widetilde{T}_{j,k}(0,\zeta,w)
=\exp\left(\zeta_{j+N(\omega)}\right)\delta_{j,k},$$
then for all $(\zeta,w)\in\Omega_{\omega}$
$$\frac{1}{\Psi(0,\zeta,w)}
=O\left(\exp\left(\alpha\omega^{\beta}\right)\right),$$
and for all $(\zeta,w)\in \mathbb{C}^{2N(\omega)}\times W$,
$\frac{\partial\widetilde{T}_{j,k}}{\partial x}(0,\zeta,w)=0$ thus
$$\frac{\partial\widetilde{\Psi}}{\partial x}(0,\zeta,w)=0.$$

At last $Q\in\Lambda_{\mathcal{Q}_{m+1}}$ being given, the following choice
of parameters
$$\zeta_j(Q)=\xi_j(Q),\;\zeta_{j+N(\omega)}(Q)
=\ln\frac{4\xi_j^2(Q)}{C_j(Q)},
\;j=1,\ldots,N(\omega),\text{ and }\;w(Q)=\omega^2Q(0),$$
is possible to get : indeed by proposition~\ref{proofestim} for all
$j=1,\ldots,N(\omega)$
$$\xi_j(Q)\text{ and }\ln\frac{4\xi_j^2(Q)}{C_j(Q)}=O(\omega^{\alpha})\,;$$
on the other hand, since $\Lambda_{\mathcal{Q}_{m+1}}$ is a compact subset
there are $c_1$ and $c_2$ such that for all
$Q\in\Lambda_{\mathcal{Q}_{m+1}}$,
$$c_1\leq Q(0)\leq c_2.$$
After choosing $r_2=2$ and setting $a_1=a=B_2(N(\omega)+1)^2$, we have
to find $B_2$ and $\varepsilon$ 
(depending on $\Lambda_{\mathcal{Q}_{m+1}}$) such that for all
$Q\in\Lambda_{\mathcal{Q}_{m+1}}$ and $\omega\geq\omega_0$ ($\omega_0$
depending on $\Lambda_{\mathcal{Q}_{m+1}}$ too),
$$\left[c_1\omega^2,c_2\omega^2\right]\subset
\left[\varepsilon B_2(N(\omega)+1)^2,
(2-\varepsilon)B_2(N(\omega)+1)^2\right].$$
Since $\varepsilon<1$ it is sufficient to get
$$
\begin{cases}
B_2\geq\frac{c_2\omega^2}{(N(\omega)+1)^2}\\
\varepsilon\leq\frac{c_1\omega^2}{B_2(N(\omega)+1)^2}.\\
\end{cases}
$$
As it was seen in the proof of theorem~\ref{optim1}, by the bounds of
Calogero the number $N_Q(\omega)$ is bounded as
$$b_1\omega\leq N_Q(\omega)\leq b_2\omega,$$
then one can choose $B_2$ and $\varepsilon$ in order to satisfy these
inequalities and the proof is finished.

\end{proof}

\begin{remark}

When $Q$ vanishes at $0$ with derivatives the
spectral measure $\sigma_{\omega}(d\tau)$ becomes
$$\sigma^0_{\omega}(d\tau)=
\begin{cases}
\frac{1}{\pi}\sqrt{\tau},\;\tau\geq0,\\
\sum_{j=1}^{N(\omega)}C_j\delta(\tau+\xi_j^2),\;\tau<0\,,
\end{cases}
$$
which gives the Gelfand-Levitan approximating formula
$Q_{\omega}^0$ which is
entirely explicit. The one obstacle is that we do not know if the
precision of approximation will still be of order of
$\frac{1}{\omega^m}$, because in order to apply the theorem~1
in~\cite{henkin}, the function $Q$ must be strictly positive on
$\mathbb{R}^+$ (in particular at $0$). However numerical
experiments make think that the approximation at order of
$\frac{1}{\omega^m}$ is valid in this case although
(see~\cite{henkin}, section~4).

\end{remark}

\subsection{Another possible application in inverse problem}

Before finishing we give here an example of inverse problem in which
we would like to use our negative results.

\begin{example}

The theorem~1.2 p. 260 in~\cite{lax-levermore} gives an original result
in the case of $L^2$-approximation : if $u$ is a negative potential of
class $C^1$ then
$\lim_{\varepsilon\rightarrow0}u(\cdot,\varepsilon)=u$ where
$$u(x,\varepsilon)=-2\varepsilon^2\frac{d^2}{dx^2}\ln\det(I+
G(x,\varepsilon)),$$
with
$$G(x,\varepsilon)=\varepsilon\left(\frac{\exp\left(-
\frac{\eta_j+\eta_k}{\varepsilon}x\right)}{\eta_j+\eta_k}C_jC_k\right)$$
and $1\leq j,\,k\leq
N(\varepsilon),\;\varepsilon=\frac{1}{\omega}$. Here we as well
consider an analytic function with respect to its eigenvalues $\eta_j$ and
characteristic values $\ln C_j$ of exponential kind. More precisely
it has the following form :
$$\widetilde{G}(x,\zeta,w)=\varepsilon\frac{\exp
(w_j+w_k)x}{w_j+w_k}\exp(\zeta_j+\zeta_k),$$
and one can choose $w_j=\frac{\eta_j}{\varepsilon^r}$, $\zeta_k=\ln C_k
j=1,\ldots,N(\varepsilon)$, with $r$ large enough such that
$\frac{\eta_j}{\varepsilon^r}$ can have a positive lower estimate. We
hope to get a result like theorems~\ref{optim1} and~\ref{optim2}.

\end{example}

\end{document}